\documentclass[prd,aps,showpacs,preprintnumbers,amsmath,amssymb,nofootinbib,showkeys, preprint]{revtex4-1}

\pdfoutput=1

\usepackage{hyperref}
\usepackage{natbib}
\usepackage{booktabs}
\usepackage{subfiles}
\usepackage{aas_macros}
\usepackage{xcolor}
\usepackage{float}

\usepackage{tikz}
\usepackage{multirow}
\usepackage{rotating}

\newcommand\tikznode[3][]%
   {\tikz[remember picture,baseline=(#2.base)]
      \node[minimum size=0pt,inner sep=0pt,#1](#2){#3};%
   }

\usepackage{graphicx}
\usepackage{dcolumn}
\usepackage{bm}
\usepackage{cases}
\usepackage[mathlines]{lineno}
\usepackage[lofdepth,lotdepth,caption=false]{subfig}

\newcommand{\unit}[1]{\ensuremath{\, \mathrm{#1}}}

\newcommand{\be}{\begin{equation}}
\newcommand{\ee}{\end{equation}}
\newcommand{\bea}{\begin{eqnarray}}
\newcommand{\eea}{\end{eqnarray}}
\newcommand{\Msun}{M_\odot}

\newcommand{\Rstar}{R_{\star}}

\newcommand{\Ncut}{N_{cutoff}}

\newcommand{\tauh}{\tau_{H}}
\newcommand{\tauhe}{\tau_{He}}
\newcommand{\vbar}{\Bar{v}}
\newcommand{\se}{\simeq}

\newcommand{\Eavg}{\langle E_R \rangle}

\begin{document}
\keywords{Dark Matter; Dark Matter capture; stars}
\title{Multi-component multiscatter capture of Dark Matter}

\author{Cosmin Ilie}
\email[E-mail at: ]{cilie@colgate.edu}
 \altaffiliation[Additional Affiliation: ]{Department of Theoretical Physics, National Institute for Physics and Nuclear Engineering,  Magurele, P.O.Box M.G. 6, Romania}
\author{Caleb Levy}
\affiliation{ Department of Physics and Astronomy, Colgate University\\
13 Oak Dr., Hamilton, NY 13346, U.S.A.
}%

\date{\today}

\begin{abstract}
In recent years, the usefulness of astrophysical objects as Dark Matter (DM) probes has become more and more evident, especially in view of null results from direct detection and particle production experiments. The potentially observable signatures of DM gravitationally trapped inside a star, or another compact astrophysical object, have been used to forecast stringent constraints on the nucleon-Dark Matter interaction cross section. Currently, the probes of interest are: at high red-shifts, Population~III stars that form in isolation, or in small numbers, in very dense DM minihalos at $z\sim 15-40$, and, in our own Milky Way, neutron stars, white dwarfs, brown dwarfs, exoplanets, etc. None of those objects are truly single-component, and, as such, capture rates calculated with the common assumption made in the literature of single-component capture, i.e. capture of DM by multiple scatterings with one single type of nucleus inside the object, are not accurate. In this paper, we present an extension of this formalism to multi-component objects and apply it to Pop~III stars, thereby investigating the role of He on the capture rates of Pop~III stars. As expected, we find that the inclusion of the heavier He nuclei leads to an enhancement of the overall capture rates,  further improving the potential of Pop~III stars as Dark Matter probes.            
\end{abstract}

\maketitle

\section{Introduction}
\label{sec:intro}

Dark Matter (DM) is one of the most longstanding mysteries of nature. It is, as of yet, an unsolved puzzle that was identified in the first half of the twentieth century. Fritz Zwicky, in 1933, coined the term \textit{Dunkle Materie} (i.e. Dark Matter) to describe the non-luminous mass that he inferred, based on data from observations of radial velocities of eight galaxies in the Coma cluster, must have been present in abundance at cluster, extra galactic scales~\cite{Zwicky:1933,zwicky1937masses}. For the next almost four decades, this idea remained highly controversial, with several important studies by Smith, Babcock, and Oort. For a review, from a historical perspective, of the early research on the question of Dark Matter see~\cite{van_den_Bergh_1999}. In the seventies, Vera Rubin and Kent Ford contributed with a major breakthrough and showed that Dark Matter must be present in abundance at galactic scales in view of the inferred ``flat'' rotation curves of stars in all the galaxies they observed~\cite{Rubin:1970}. In the subsequent half century since then, the Dark Matter hypothesis became supported by more and more experimental evidence. For a review, see~\cite{Freese:2017dm}. The consensus that emerged, from various observations that are very different in nature, is that only $\sim20\%$  of the matter in the universe is made of regular, baryonic matter. The other $\sim80\%$ is Dark Matter, which manifests itself via its gravitational effects at various scales. In addition, DM leaves its imprint in the Cosmic Microwave Background (CMB) radiation \cite{Komatsu:2009,Komatsu:2011,Ade:2015,Aghanim:2018}, and this leads to some of the most precise measurements of the amount of Dark Matter in the Universe. This phenomenon can be understood in view of the acoustic oscillations of the photon-baryon plasma in the early universe, that are driven by the restoring gravitational force provided by the DM potential wells of the over-dense regions that form by the growth of the seeds left over from anisotropies in the inflation field. 

This gravity-driven growth continues, in a hierarchical fashion, ultimately leading to the formation of the large structures we see today in the universe, such as galaxy clusters. Dark Matter forms minihalos that eventually merge and grow larger and more and more massive, and have a rich sub-structure. 
Typically, those over-dense regions of the universe, dominated by DM, are connected by DM filaments, as shown in numerous numerical simulations. As such, DM provides the gravitational well that attracts regular, baryonic matter, which eventually collapses to form galaxies and galaxy clusters. Gravitational lensing was used to confirm the abundance of DM at galactic scales with the SDSS survey~\cite{AdelmanMcCarthy:2005}, and to map the structures DM forms at galaxy cluster~\cite{Natarajan:2017} and cosmological~\cite{Madhavacheril:2014,Vikram:2015,Hikage:2018} scales.

DM Direct detection experiments are extremely challenging. They are very sensitive, to the point of being able to detect the minute amount of energy a Dark Matter particle deposits inside the detector as it collides with an atomic nucleus~\cite{Goodman:1985,Drukier:1986}. Shielding from cosmic ray backgrounds means that these experiments have to be performed in deep, underground laboratories. Of the ten currently operational direct detection experiments, only the DAMA/LIBRA experiment in Gran Sasso, Italy produced a  detection signal~\cite{BERNABEI:1998,Bernabei:2014,Bernabei:2018}. Since 1998, the DAMA/LIBRA experiment finds an annual modulation in its signal that matches the modulation predicted by~\cite{Drukier:1986}. Although this is the cleanest hint of a Dark Matter detection yet, unfortunately, it has not been confirmed by other direct detection experiments exploring the same region of the parameter space, such as XENON1T. To settle this controversy, a new NaI experiment (the same detector material as DAMA/LIBRA) has been developed: COSINE~\cite{Adhikari:2018}. It will soon either refute or confirm the DAMA signal.~\footnote{Recently, another experiment (ANAIS) has analysed their three year data and found no annual modulation~\cite{Amare:2021}.} Another hint of DM detection came recently from XENON1T, the world's most sensitive DM direct detection experiment. An excess in the electronic recoil events could be explained by, among other things, solar axions~\cite{Aprile:2020}. While solar axions are not a Dark Matter candidate, their detection, if confirmed, would be the first discovery of a particle outside of the standard model of particle physics. This would provide insights into the production of axions in the early universe, which could serve as Dark Matter candidates. 

In lack of clear, independently confirmed detection signals from direct detection experiments, we are left with exclusion limits on how strongly DM and baryonic matter can interact. As experiments become more and more sensitive, they rule out larger and larger swaths of the possible DM-nucleon scattering cross section $\sigma$, vs DM particle mass $m_X$, parameter space. However, an increase in sensitivity comes at a price. In the near future, it is expected that the XENON1T experiment will become sensitive to neutrinos~\cite{Billard:2013}. At that stage, any possible DM signal would be swamped by an overwhelming neutrino background, the so-called neutrino floor. As such, new detection strategies will have to be implemented. In this paper, we discuss and further demonstrate the value of one such strategy, which relies on the capture of Dark Matter by the first generation of stars, the so called Population~III (Pop~III) stars.

The importance of astrophysical objects as probes of Dark Matter has been long recognized in the literature. For example, the first seminal papers on capture of Dark Matter all deal with the potentially observable effects of Dark Matter (DM) captured by our Sun~\cite{Press:1985,Spergel:1985,Gould:1987,Gould:1987resonant} and the Earth~\cite{Gould:1992ApJ}. Simply put, Dark Matter particles within the Dark Matter halos surrounding any galaxy, have the potential to be slowed down by collisions with nuclei inside the dense environments of stars, or other compact astrophysical objects. Once they are slowed below the escape velocity, those particles become gravitationally trapped by the captor object and, eventually, sink in towards their center, where they could annihilate efficiently and produce heat with observable effects. The formalism for calculating the capture rates was initially limited by the requirement that, on average, the Dark Matter particles, as they cross the capturing object, will experience at most one collision, i.e. single scatter capture~\cite{Press:1985,Spergel:1985,Gould:1987,Gould:1987resonant}. This was extended to the case of finite optical depth and used to study WIMP capture by the Earth~\cite{Gould:1992ApJ}. 

Recently, the use of multiscatter capture has re-emerged in the literature to calculate capture rates in very dense environments~\cite{Bramante:2017,Dasgupta:2019juq,Ilie:2020Comment,Bell:2020}, such as neutron stars, or in the high cross section limit, where, on average, a DM particle will collide multiple times per crossing with targets inside the object. Based on the potentially observable effects due to captured Dark Matter, several classes of objects have been investigated as useful probes of DM. Below we include a non-exhaustive list of the more recent papers where such effects have been analysed for: Pop~III stars~\cite{Freese:2008cap,Iocco:2008,Ilie:2019,Ilie:2020PopIIIa,Ilie:2020BNFa}, Neutron Stars~\cite{Baryakhtar:2017,Bramante:2017,Raj:2017wrv,Croon:2017zcu,Bell:2018pkk,Chen:2018ohx,Gresham:2018rqo,Acevedo:2019,Bell:2019pyc,Hamaguchi:2019oev,Leroy:2019,Leung:2019,Joglekar:2019,Bell:2020,Bell:2020b,Bell:2020NSSINS,Garani:2020,Genolini:2020,Joglekar:2020,Keung:2020,Kumar:2020,Perez-Garcia:2020}, White Dwarfs~\cite{Bertolami:2014wua,Bramante:2017,Dasgupta:2019juq,Horowitz:2020axx,Panotopoulos:2020kuo}, and exoplanets~\cite{Leane:2020wob}. The capture mechanism in most of those papers is commonly assumed to be via collisions with one unique nucleus, or, in the case of neutron stars, with neutrons. We point out that this is not a valid assumption as none of these objects are purely  single-component. Thus, for all astrophysical objects considered as probes of Dark Matter, one needs to include the subtleties of having various target nuclei that could slow down the DM particles. The aim of this paper is to provide a framework for such calculations. 

Our paper is organized as follows: In Sec.~\ref{sec:TwoCompformalism}, we extend the single-component multiscatter formalism of~\cite{Gould:1992ApJ,Bramante:2017} to a more general, two-component multiscatter scenario. In Sec.~\ref{sec:MCMSPopIIIStars}, we validate our formalism by applying it to Pop~III stars, and recover, in the appropriate limits (Hydrogen fraction equal 1 or 0), the results we would expect from the single-component multiscattering formalism. In Sec.~\ref{sec:MaxMstar}, we obtain projected upper bounds on stellar Pop~III masses, in view of the possible captured DM annihilations. In Sec.~\ref{sec:BoundsOnDM}, we refine the forecasted bounds on the proton-DM scattering cross section previously obtained by using Pop~III stars in~\cite{Ilie:2020BNFa,Ilie:2020PopIIIa} by the inclusion of the effects due to the He nuclei on the capture of Dark Matter. We end with Sec.~\ref{sec:Conclusions}, where we summarize our results and present conclusions. We include the following appendices: Appendix~\ref{sec:SigmaAppendix} presents the standard way to estimate how the cross section of interaction between He nuclei and DM relates to the proton-DM interaction cross section. In Appendix~\ref{section:AnalyticExpressions}, we derive and validate useful analytic approximations of the total capture rate of DM by a two-component object, such as a Pop~III star. In Appendix~\ref{sec:ApMCMS}, we present an extension of the two-component DM capture formalism to an arbitrary number of components. Finally, in Appendix~\ref{Sec:NumericalTest}, we test the validity of our chosen numerical convergence criteria, used throughout our paper to calculate numerically the DM capture rates in Pop~III stars, and discuss the computational implications of the multi-component formalism.


\section{Two-component Multiscattering Formalism}\label{sec:TwoCompformalism}

The multiscatter formalism for a single-component astrophysical object was developed in \cite{Gould:1992ApJ,Bramante:2017,Ilie:2020Comment} and has subsequently been used to calculate the effects of Dark Matter capture on various astrophysical bodies, including neutron stars, white dwarfs, Pop~III stars, and exoplanets
\cite{Bramante:2017,Ilie:2019,Ilie:2019Erratum,Ilie:2020Comment,Ilie:2020PopIIIa,Ilie:2020BNFa,Leane:2020wob}. In this paper, we present a method for calculating the DM capture rate in objects composed of more than one element, then apply this formalism to Pop~III stars. This more general formalism allows one to account for the varying composition of astrophysical bodies in the calculation of DM capture rates. For the purposes of presenting the two-component formalism most generally, we will focus on an arbitrary two-component object composed of nuclei A and nuclei B with mass fractions given by: $f_A \equiv \frac{M_A}{M_{total}}$ and $f_B \equiv \frac{M_B}{M_{total}}$, where $M_{total}$ is the total mass of the object. The schematic differential capture rate for DM particles by an object is
given by \cite{Bramante:2017}:
\be\label{eq:captureSchem}
    \frac{dC}{dV d^3u} = dF(n_X, u, v_{obj}, v_{esc}^{halo})\Omega(n_T(r), w(r), \sigma, m_n, m_X),
\ee
where $dF$ is the differential flux of DM particles, $\Omega$
is the probability of a given DM particle becoming gravitationally
bound after collisions with the object's constituents, $n_X$ is the local DM number density, $u$ is the velocity of DM particles far from the object,
$v_{obj}$ is the velocity of the object relative to the DM halo,
$v_{esc}^{halo}$ is the escape velocity of the DM halo, $n_{T}(r)$ is the
number density of the object's constituents at a radius r from the center of the
object, and $w(r)^2 = u^2 + v_{esc}(r)^2$, with $v_{esc}(r)^2 \equiv \frac{2 G
M_{obj}(r)}{r}$. We make the following assumptions for the purposes of this method: no relative velocity between the object and the thermal DM distribution ($v_{obj} \to 0$), infinite escape speed for the DM halo ($v_{esc}^{halo} \to \infty$), a uniform density object in both components, and a constant escape velocity throughout the object ($v_{esc}(r) = v_{esc}(R_{obj})$).

We now go on to estimate the error of making the simplifying assumptions above in the context of Pop~III stars.~Firstly, the assumption of zero relative velocity between the object and halo ($v_{obj} \to 0$) is addressed in Ref.~\cite{Ilie:2021vel}, where an analytic approximation of the suppression of the capture rates whenever $v_{obj}\neq 0$ is given. In this reference, it is shown that a realistic relative velocity for Pop~III stars leads to a negligible (less than an order of magnitude) suppression of the DM capture rates and can thus be safely ignored in this context. The assumption of infinite halo escape velocity is a common one made in the literature (See \cite{Gould:1992ApJ,Bramante:2017}). To check whether this approximation holds in this context, we estimated the halo escape velocity for an adiabatically-contracted DM microhalo, likely candidates for hosts of Pop~III stars \cite{Freese:2008dmdens}, and compared it to the average dispersion velocity of DM particles. Comparing these quantities provides a good estimate since this assumption has implications for the Maxwell-Boltzmann velocity distribution used in this paper which has an exponential term of the form $\text{Exp}\left[-\frac{3 u^2}{2 \vbar^2}\right]$. We find that the value of the ratio $(v_{esc}^{\text{halo}}/\vbar)^2$ is $\sim 10$ in the region from which DM particles are captured, which is sufficient to take $v_{esc}^{\text{halo}} \to \infty$ since DM particles with this velocity or higher would be exponentially suppressed by at least a factor of $\sim e^{-10}$. Moreover, the probability that a DM particle has a velocity larger than the DM halo escape velocity is less than one in sixty million, further reinforcing the validity of our assumption: $v_{esc}^{\text{halo}} \to \infty$. To estimate the error of assuming a uniform density object on the capture rate of Pop III stars, consider the differential optical depth from a small portion of a path $\ell$ which a DM particle may take through the star: $d\tau = n_T(r) \sigma d\ell$. Integrating over the path would then give the average number of scatters a DM particle would undergo along that path. To estimate the effect of taking $n_T(r) \simeq n_{avg}$, one may consider a path passing close by the center of the star with a functional approximation of $n_T(r)$. To do this, we use the $n=3$ polytropic assumption, valid whenever the ratio between the
radiation pressure and the gas pressure is a constant throughout the star, as is the case for the radiation
pressure dominated $M_\star \gtrsim 100 M_\odot$ Pop~III stars on the zero age main sequence. In this model, the density profile for each component is given by: $n_T(r) = n_c \theta^3(r)$, where $n_c$ is the central density, and $\theta$ is the Lane-Emden function for $n=3$. We find that integrating along straight-line paths close to the center of the star ($\sim 0.1 R_\star$) leads to an enhancement of an order of magnitude of the optical depth, which is suggestive of higher DM capture rates. Thus, the capture rates and projected bounds we present in this paper are conservative for this reason. Using the polytropic model, one can also check the assumption of constant escape velocity, $v_{esc}(r) \simeq v_{esc}(R_\star)$. Given the polytropic density profile above, $n_T(r) = n_c \theta^3(r)$, the mass enclosed at a given radius $M(r)$ can be found by integrating the density from the center of the star to the desired radius. One can then find the escape velocity at a given radius with the following prescription: $\frac{1}{2} v_{esc}^2(r) = \int_{r}^{\infty} \frac{G M(r')}{(r')^2} dr'$. Applying this to Pop~III stars, we find that the escape velocity is enhanced by a maximum factor of $\sim 2$. This leads to a maximum enhancement of the total capture rate within an order of magnitude, which can be seen by the relationship $C_{tot} \sim v_{esc}^{4}$ derived in the analytic capture rates in Eq.~\ref{equation:CtotApprox_Pop3}, thus proving further that our estimates are conservative.

The total capture rate of a DM particle of mass $m_X$ is:
\be
    C_{tot}(m_X) = \sum_{N = 1}^{\infty} C_{N},
    \label{equation:CtotBasic}
\ee
where $C_{N}$ is the capture rate after exactly N scatters. The optical
depth for a two-component object can be represented by two separate parameters, $\tau_A \equiv 2 R_{obj} \sigma_A n_A$ and $\tau_B \equiv
2 R_{obj} \sigma_B n_B$, where $n_A$ ($n_B$) is the number density of nucleus A (B) and $\sigma_A$ ($\sigma_B$) is the
DM scattering cross section with nucleus A (B). The
optical depths are defined to represent the average number of
 scatters per crossing a DM particle has with the corresponding component, i.e., $\langle N_A\rangle \approx \tau_A$ and $\langle N_B\rangle \approx \tau_B$. The DM scattering cross section for interactions with nucleus A and nucleus B are ultimately a function of the spin-independent (SI) or spin-dependent (SD) elastic DM-nucleon cross sections (See Appendix \ref{sec:SigmaAppendix} for a review on DM-nucleon scattering cross sections). 
 For a DM particle undergoing multiple scatters in a single-component object,
the partial capture rate $C_N$, is calculated by taking into account the
flux of DM particles on the object and two probabilistic parameters:
$p_N(\tau)$ and $g_N(w)$, where $p_N(\tau)$ is the probability of a DM
particle undergoing exactly N scatters while traversing the object and
$g_N(w)$ is the probability of the DM particle falling below the object's
escape velocity after N scatters. It is important to note that the formalism separates the probability for the average number of scattering events that will occur, which depends only on the cross section and the capturing object, from the probability that a given number of scatterings will lead to capture, which depends on the kinematics of the collisions. First, we will discuss the effect of a
multi-component object on $p_N(\tau)$. 

The probability that a DM particle undergoes N scatters is given by a Poisson distribution modified to factor all the possible incidence angles of DM particles on the object \cite{Bramante:2017,Ilie:2020Comment}:
\be
p_N(\tau) = 2\int_0^1\text{Poisson}(y\tau,~N)~y  dy= (2/N!) \int_{0}^{1} y e^{-y \tau}(y \tau)^N dy,
\ee
where $y \equiv \cos\alpha$ and $\alpha$ is the incidence angle of the DM particle on the object. To show how this equation arises, consider the differential flux on a spherical surface of radius $R_a$ far from the star's gravitational potential for DM particles within a small incidence angle $\theta$: $dF = 4\pi R_a^2 \times \frac{1}{2} f(u) \vec{u}\cdot\hat{R_a} du~ d(\cos\theta) = 4\pi R_a^2 \times \frac{1}{2} f(u) u \cos\theta du~ d(\cos\theta)$, where $\vec{u}$ is the initial DM velocity, and $\hat{R_a}$ is a vector normal to the sphere. Conservation of angular momentum requires that $u R_a \sin\theta = w R_{obj} \sin\alpha$, where $w$ is the DM velocity at the object, and $\alpha$ is the incidence angle of the DM particle on the object. One can then rewrite the differential flux as: $dF = \pi R_{obj}^2 \times \frac{f(u)}{u} w^2 du~d(\cos^2\alpha)$. Then, for a DM particle with an incidence angle $\alpha$ on the object, the straight line path through the object will lead to an average number of scatters given by $\tilde{\tau} = \tau \cos\alpha$, and thus a probability for $N$ scatters of the form $\tilde{p}_N(\tau,~\alpha) = \text{Poisson}(\tau\cos\alpha,~N)$. The factor $d(\cos^2\alpha)$ from the differential flux then becomes absorbed into the probability function and integrated over all possible incidence angles of DM on the star, giving: $p_N(\tau) = 2 \int_0^1 \text{Poisson}(y\tau, N)~y dy$, where $y\equiv \cos\alpha$. It was shown in  Ref.~\cite{Ilie:2019} that $p_N(\tau)$ has the following closed form:
\be\label{eq:pn_tau}
    p_{N}(\tau) = \frac{2}{\tau^2} \left(N + 1 - \frac{\Gamma(N + 2, \tau)}{N!}\right).
\ee
In order to extend this to a two-component object, we must
consider, generally, the possibility of scatters with nuclei A (denoted by $i$) and
scatters with nuclei B (denoted by $j$) where, $N = i + j$ is the total number of
scatters a DM particle may undergo. Following this, we define two
probability functions:
\begin{equation}
    p_{i}(\tau_{A}) = 
    \begin{cases}
        \frac{2}{\tau_{A}^2} \left(i + 1 - \frac{\Gamma(i + 2, \tau_{A})}{i!}\right), & \text{if } \tau_{A} > 0\\
        \Theta(-i), & \text{if } \tau_{A} = 0\\
    \end{cases}
    \label{equation:pntau1}
\end{equation}
\begin{equation}
    p_{j}(\tau_{B}) = 
    \begin{cases}
        \frac{2}{\tau_{B}^2} \left(j + 1 - \frac{\Gamma(j + 2, \tau_{B})}{j!}\right), & \text{if } \tau_{B} > 0\\
        \Theta(-j), & \text{if } \tau_{B} = 0,\\
    \end{cases}
    \label{equation:pntau2}
\end{equation}
where $p_i(\tau_A)$ is the probability of undergoing i scatters with nuclei A
and $p_j(\tau_B)$ is the probability of undergoing $j$ scatters with
nuclei B. The heaviside step function, $\Theta$, is introduced for
consistency with the single-component multiscatter formalism. It is defined
such that when $\tau = 0$, the probability of undergoing 0 scatters is 1 and
the probability of undergoing $N \geq 1$ scatters is 0. Physically, this
means that if there does not exist a given component in an object (i.e $\tau =
0$), there is a zero probability of colliding with it. We will next consider the
probability of capture after N scatters $g_N(w)$.

A DM particle incident on the object will have an initial kinetic energy $E_0
= \frac{1}{2} m_X w^2$. After elastically colliding with a nucleus, the particle will lose energy defined by the kinematic equation
$\Delta E = z \beta_+ E_0$, where $z \in [0,1]$ is related the scattering angle in the center of mass frame by $z=\sin^2(\theta_{\text{CM}}/2)$ \cite{Dasgupta:2019juq}, and $\beta_{\pm} = \frac{4 m
m_X}{(m_X \pm m)^2}$. Following this, the kinetic energy after one
collision is $E_k = (1 - z_k \beta_+) E_{k-1}$, and the corresponding DM
velocity becomes $v_k = (1 - z_k \beta_+)^{1/2} v_{k-1}$.
In a multi-component object, the energy lost in a given collision depends on
which constituent the DM particle collides with. For this reason, we define
$\beta_{\pm}^{A} = \frac{4 m_A m_X}{(m_X \pm m_A)^2}$ and
$\beta_{\pm}^{B} = \frac{4 m_{B} m_X}{(m_X \pm m_{B})^2}$ for
collisions with nuclei A and B respectively. After exactly $N = i + j$
scatters, the kinetic energy and velocity
of the DM particle becomes:
\begin{equation}
    E_{ij} = E_0 \prod_{k = 1}^{i}(1 - z_{k} \beta_{+}^{A}) \prod_{n = i+1}^{i+j}(1 - z_{n} \beta_{+}^{B}),
    \label{equation:E}
\end{equation}
\begin{equation}
    v_{ij} = w \prod_{k = 1}^{i}(1 - z_{k} \beta_{+}^{A})^{\frac{1}{2}} \prod_{n = i+1}^{i+j}(1 - z_{n} \beta_{+}^{B})^{\frac{1}{2}}.
    \label{equation:v}
\end{equation}
Taking the capture condition as
$v_{ij} < v_{esc}$ and integrating over all possible paths a DM
particle can take through the object while undergoing i scatters with nucleus A and j scatters with nucleus B, we define $g_{ij}(w)$:
\be\label{eq:gij_integral}
    g_{ij}(w) = \int_{0}^{1} dz_{1} \int_{0}^{1} dz_2 \cdots \int_{0}^{1} dz_{i} \int_{0}^{1} dz_{i+1} \cdots  \int_{0}^{1} dz_{i+j} \Theta\left(v_{esc} - v_{ij}\right),
\ee
where $z_k$ is integrated until $k = i$, then $z_n$ from $n = i + 1$ to $N = i+j$, representing the scatters with nucleus A and B respectively. In this integral, the $\Theta$ function describes the probability
of capture of a DM particle that traces a path through the object described by
$i$ collisions with nucleus A and $j$ collisions with nucleus
B. Defining $g_{ij}$ in this way is done to ask the following question: if a specific DM particle with velocity $w$ collides $i$ times with A and $j$ times with B, will it's velocity fall below the star's escape velocity? This provides a way to separate the kinematics, which depend on the DM mass and velocity, from the average number of collisions a DM particle will actually undergo, which depends only on the cross section and the capturing object. Ultimately, this allows one to find the regions in the DM's initial velocity space that could be captured for a given combination of collisions, and what the corresponding rate would be. We can simplify Eq.~(\ref{eq:gij_integral}) further by assuming that there is no preferred scattering direction, as done in the single-component multiscatter formalism \cite{Bramante:2017}, thereby taking the average value for $z_{k}$ and $z_{n}$. Appendix~\ref{sec:SigmaAppendix} contains details on estimating $\langle z\rangle$. The probability of capture after $i$ scatters with nucleus A and $j$ scatters with nucleus B then becomes:
\begin{equation}\label{eq:gij_solved}
    g_{ij}(w) = \Theta\left(v_{esc}\prod_{k = 1}^{i}\left(1 - \langle z_{A}\rangle \beta_{+}^{A}\right)^{-\frac{1}{2}} \prod_{n = 1}^{j}\left(1 - \langle z_{B}\rangle \beta_{+}^{B}\right)^{-\frac{1}{2}} - w\right).
\end{equation}

In a single-component context, the partial capture rate $C_N$, can be calculated by multiplying the rate at which DM particles pass through the object with the probability of being captured after N scatters $p_N(\tau) g_N(w)$. After exactly N collisions, the capture rate is given by the following phase-space integral:
\begin{equation}
    C_{N} = \pi R^{2} p_{N}(\tau) \int_{v_{esc}}^{\infty} dw \frac{f(u)}{u^2} w^3 g_{N}(w),
    \label{equation:CnSingle}
\end{equation}
where $f(u)$ is the DM velocity distribution. In a multi-component context, calculating the partial capture rate involves summing over the capture rates associated with all possible combinations of scattering events with nucleus A and nucleus B. For a given number of scatters $N$, there exists $N + 1$ possible combinations of scattering events with nucleus A and nucleus B. Substituting $p_N(\tau) g_N(w)$ with $p_i(\tau_{A}) p_j(\tau_B) g_{ij}(w)$, the probability of capture after exactly $i$ scatters with nucleus A and $j$ scatters with nucleus B, and summing over all $N+1$ possible ways this could happen gives:
\begin{equation}
    C_{N} = \sum_{i = 0}^{N}\left[\pi R^{2} p_{i}(\tau_{A}) p_{j}(\tau_{B}) \int_{v_{esc}}^{\infty} dw \frac{f(u)}{u^2} w^3 g_{ij}(w)\right],
    \label{equation:CnMulti}
\end{equation}
where we remind the reader that $j = N-i$. Analytically evaluating this integral under the assumption of a maxwellian velocity distribution with an average speed of $\vbar$ gives:
\begin{equation}
    C_{N} = \sum_{i = 0}^{N}\frac{\pi}{3} R^2 p_{i}(\tau_{A}) p_{j}(\tau_{B}) \frac{\sqrt{6} n_X}{\sqrt{\pi} \Bar{v}}\left((2 \Bar{v}^2 + 3 v_{esc}^2) - (2 \Bar{v}^2 + 3 v_{ij}^2) \text{exp}\left(-\frac{3(v_{ij}^2 - v_{esc}^2)}{2 \Bar{v}^2}\right)\right),
    \label{equation:CnSolved}
\end{equation}
where:
\begin{equation}
    v_{ij} \equiv v_{esc} \left(1 - \langle z_{A}\rangle \beta_{+}^{A}\right)^{-\frac{i}{2}} \left(1 - \langle z_{B}\rangle \beta_{+}^{B}\right)^{-\frac{j}{2}}.
    \label{equation:vnk}
\end{equation}
Equation~\ref{equation:CnSolved} arises from imposing a cutoff on the integral over velocity which arises from the step function in Eq.~(\ref{eq:gij_solved}), the probability for capture after $i$ scatters with A and $j$ scatters with B. This is because for a given scattering scenario defined by $i$ scatters with A and $j$ scatters with B, there exists a velocity above which DM will not be captured, given by the first term in the step function of Eq.~(\ref{eq:gij_solved}).

We now go on to analytically show how the multi-component multiscatter formalism reduces to the single-component multiscatter formalism in the appropriate limit. First, we consider the case of an object made entirely of nucleus A, i.e., $f_A = 1$ and $f_B = 0$. The single-component formalism would describe the capture rate of this object as \cite{Bramante:2017,Ilie:2020Comment} :
\begin{equation}
    C_{N} = \frac{\pi}{3} R^2 p_{N}(\tau_{A}) \frac{\sqrt{6} n_X}{\sqrt{\pi} \Bar{v}}\left((2 \Bar{v}^2 + 3 v_{esc}^2) - (2 \Bar{v}^2 + 3 v_{N}^2) \text{exp}\left(-\frac{3(v_{N}^2 - v_{esc}^2)}{2 \Bar{v}^2}\right)\right),
    \label{equation:CnSCMS}
\end{equation}
with:
\be
v_N = v_{esc} (1 - \langle z_A \rangle \beta_+^A)^{-\frac{N}{2}}.
\ee
In the multi-component multiscatter formalism, we first point out that in the limit of $f_A = 1$, the probability functions $p_i$ and $p_j$ become:
\begin{equation}
    p_{i}(\tau_{A}) = 
        \frac{2}{\tau_{A}^2} \left(i + 1 - \frac{\Gamma(i + 2, \tau_{A})}{i!}\right),
\end{equation}
\be
    p_{j}(\tau_B = 0) = 
    \begin{cases}
        1, & \text{if } j = 0\\
        0, & \text{if } j > 0.\\
    \end{cases}
    \label{equation:pktau0}
\ee
This means that the partial capture $C_N$, will simplify as all the terms in the sum where $j > 0$ ($i \neq N$) equal 0, leaving only the last term (i = N). Keeping only the $i = N$ term, we also see the following simplification (recalling that when $i = N$, $j = 0$):
\be
v_{N,0} = v_N = v_{esc} \left (1 - \langle z_A \rangle \beta_+^A\right)^{-\frac{N}{2}}.
\ee
It is now simple to show that Eq.~(\ref{equation:CnSolved}) reduces exactly to Eq.~(\ref{equation:CnSCMS}), verifying analytically the multi-component formalism reduces to the single-component formalism in the appropriate limit. Also, note the symmetry between A and B in the multi-component multiscatter formalism, meaning this process is exactly equivalent for $f_B = 1$. In Sec.~\ref{sec:MCMSPopIIIStars}, we present a numerical verification of this reduction for Pop~III stars. 

In this section, we presented a method for calculating DM capture rates in two-component astrophysical objects. We ended by verifying that this method analytically collapses to the well established single-component formalism. In the next section, we go on to apply this formalism to the first stars to investigate the effects of helium on DM capture rates.

\section{Multi-component Capture in Pop~III Stars}\label{sec:MCMSPopIIIStars}

In this section, we calculate upper bounds on DM capture rates and DM luminosity from DM-DM annihilation for Pop~III stars composed of $\sim 25\%$ helium and $\sim 75\%$ hydrogen, applying the multi-component multiscatter formalism. We compare our results to the previously used approximation of a pure-hydrogen Pop~III star to demonstrate the effects of helium on the capture process. Pop~III stars are believed to have been formed at or around the center of dense DM mini-halos out of pristine gas from big bang nucleosynthesis (BBN) at redshifts $z = 10-50$~\cite{Bromm:1999, Bromm:2003,Bromm:2009}. As they form from the collapse of primordial gas, their composition at formation is well approximated by the mass fractions predicted by BBN, i.e. $25\%$ helium and $75\%$ hydrogen~\cite{Cyburt:2016}. For this reason, we have adopted the following mass fractions for zero-age main sequence (ZAMS) Pop~III stars: $f_{He} = 0.25$ and $f_{H} = 0.75$. Note that in reality, these mass fractions will vary as the star burns hydrogen, however, this makes capture rates conservative as higher helium fractions from hydrogen burning will lead to higher capture rates. In addition, we have assumed a standard adiabatically contracted NFW density profile for the DM mini-halos in which Pop~III stars formed. This profile is well established in the literature and has been shown to agree excellently with simulation data in the inner-parsec region of these mini-haloes \cite{Abel:2001, Freese:2008dmdens} (See Appendix E of \cite{Ilie:2020PopIIIa} for more details on this assumption).~Following this, we get a range of values for the halo mass, $M_{halo} = 10^5-10^6 M_\odot$, DM dispersion velocity, $\vbar = 1 - 15$ km/s, and the density at the center of the halo, $\rho_X = 10^{13} - 10
^{16}$ GeV/cm$^3$ of which we adopt the following fiducial values:
\be
M_{halo} = 10^6 M_\odot,
\ee
\be
    \vbar = 10 \text{ km/s},
\ee
\be
    \rho_X = 10^{14} \text{ GeV/cm$^3$}.
\ee
Note that the capture rates scale linearly with $\rho_X$, so it is straightforward to adjust our results for any other assumption made on $\rho_X$. 
To calculate DM capture rates, we require the DM-nucleus scattering cross section. In Appendix \ref{sec:SigmaAppendix}, we provide a brief review of the theory behind DM-nucleon scattering and show how we obtain the following expressions in the context of Pop~III stars:
\be\label{eq:sigmaH}
\sigma \equiv \sigma_H = \sigma_0^{SI-p},
\ee
\be\label{eq:sigmaHe}
\sigma_{He} = 
        4^4 \sigma_0^{SI-p} \langle F^2(E_R)\rangle,
\ee
where $\sigma_H$ is the DM-hydrogen cross section, $\sigma_{He}$ is the DM-helium cross section, $\sigma_0^{SI-p}$ is the ``standard" spin-independent (SI) DM-proton cross section in the $q \to 0$ limit, and $\langle F^2(E_R)\rangle$ is the average of the helm form factor across all recoil energies. Note the null contribution of Helium to spin-dependent (SD) interactions due to its 0 spin. For protons, in reality, the total cross section is a sum of the SI and SD cross sections. However, following the practice of direct detection experiments, one can make the assumption that one dominates over the other. We make this simplifying assumption for consistency, since we are contrasting the projected bounds obtained from our method to those of direct detection experiments. In this paper, we assume SI interactions dominate in order to study DM capture in a multi-component context, as both hydrogen and helium interact via the SI channel. If one assumes that the SD cross section is dominant, capture in Pop~III stars is relevant only for DM-proton collisions and is thus more suitable for the single-component multiscatter formalism, as done in \cite{Ilie:2019,Ilie:2019Erratum,Ilie:2020PopIIIa,Ilie:2020BNFa}. Also, note the significant enhancement of the cross section for helium interactions due to the higher mass of helium nuclei, as well as the minor suppression from the nuclear form factor. For the purposes of this paper, one initial goal we have is to place projected upper bounds on Pop~III stellar masses by using the XENON1T bounds on SI interactions \cite{Aprile:2018} for DM masses $\gtrsim 10^2$ GeV, which scale linearly as a function of DM mass in the following way:
\begin{equation}\label{Eq:X1Tbounds}
    \sigma_0^{SI-p}\lesssim 8\times 10^{-47}~\unit{cm}^2\left(\frac{m_X}{10^2~\unit{GeV}}\right).
\end{equation}
This linear relationship can be understood from the fact that the total nuclear recoil rates in direct detection experiments scales like the DM flux, whenever the DM particle mass is significantly higher than the target nuclei. In turn, the flux scales like $\propto n_X^{local} \sigma$. The local mass density of DM, $\rho_X^{local}=n_X^{local}/m_X$, is a well-constrained parameter, and thus increasing DM mass leads to a lower flux of dark matter particles incident on detectors and hence a linear dependence on mass for the upper bounds on $\sigma$ placed by direct detection. For Pop~III stellar parameters, we rely on numerical simulations from \cite{Windhorst:2018}. These parameters can be found in Table \ref{table:StellarData} along with the star's surface escape velocity $v_{esc}$.
\begin{table}[ht]
\centering
\begin{tabular}{c c c c c c}

\hline\\
$M_\star [M_\odot]$    & $R_\star [R_\odot]$     & $v_{esc} [\text{cm} s^{-1}]$     & $L_\star [L_\odot]$    \\ \\ \hline
1    & 0.875 & $0.660 \times 10^8$ & $1.91 \times 10^0$ \\
1.5  & 0.954 & $0.774 \times 10^8$ & $1.05 \times 10^1$ \\
2    & 1.025 & $0.862 \times 10^8$ & $3.29 \times  10^1$ \\
3    & 1.119 & $1.011 \times 10^8$ & $1.46 \times 10^2$ \\
5    & 1.233 & $1.243 \times 10^8$ & $8.46 \times 10^2$ \\
10   & 1.400 & $1.650 \times 10^8$ & $7.27 \times 10^3$ \\
15   & 1.515 & $1.943 \times 10^8$ & $2.34 \times 10^4$ \\
20   & 1.653 & $2.148 \times 10^8$ & $5.11 \times 10^4$ \\
30   & 2.123 & $2.321 \times 10^8$ & $1.45 \times 10^5$ \\
50   & 2.864 & $2.580 \times 10^8$ & $4.25 \times 10^5$ \\
100  & 4.118 & $3.043 \times 10^8$ & $1.40 \times 10^6$ \\
300  & 7.408 & $3.930 \times 10^8$ & $6.57 \times 10^6$ \\
1000 & 12.85 & $5.447 \times 10^8$ & $2.02 \times 10^7$ \\ \hline
\end{tabular}

\caption{Stellar mass $M_\star$, radius $R_\star$, surface escape velocity $v_{esc}$, and luminosity $L_\star$, for simulated Pop~III stars. These parameters are used to calculate projected DM capture rates in these objects.}
\label{table:StellarData}
\end{table}
In order to calculate the total capture rate numerically, we have set a condition for the infinite sum given by Eq.~(\ref{equation:CtotBasic}) to be truncated at some $N_{cutoff}$ when the sum has converged. This is possible in light of the fact that the partial capture rate $C_N$, is rapidly driven to 0 after the number of collisions surpasses the average number of collisions $\tau$. We have adopted the following cutoff conditions to numerically calculate the capture rate in Pop~III stars:
\be\label{eq:Cutoff1}
    \left| (C_{tot, N_{cutoff}}/C_{tot, N_{cutoff}-1}) - 1 \right| \leq 0.0001,
\ee
\be\label{eq:Cutoff2}
    C_{N_{cutoff}} < C_{N_{cutoff} + 1}.
\ee
This reduces the total capture rate from an infinite sum to a partial sum:
\be\label{eq:Ctot_toNcut}
    C_{tot} \approx \sum_{N = 1}^{N_{cutoff}} C_N.
\ee
In general, the value of $N_{cutoff}$ depends proportionally on the sum of optical depths, $\sum_i \tau_i$. Thus, one can estimate the number of terms that will be required to sum and find typical values of $N_{cutoff}$ by calculating $\sum_i \tau_i$. See Fig.~\ref{fig:Ncut_sumTau} for verification of this cutoff criteria.

\begin{figure}   [bht]
\centering
\includegraphics[angle=0,scale=0.45]{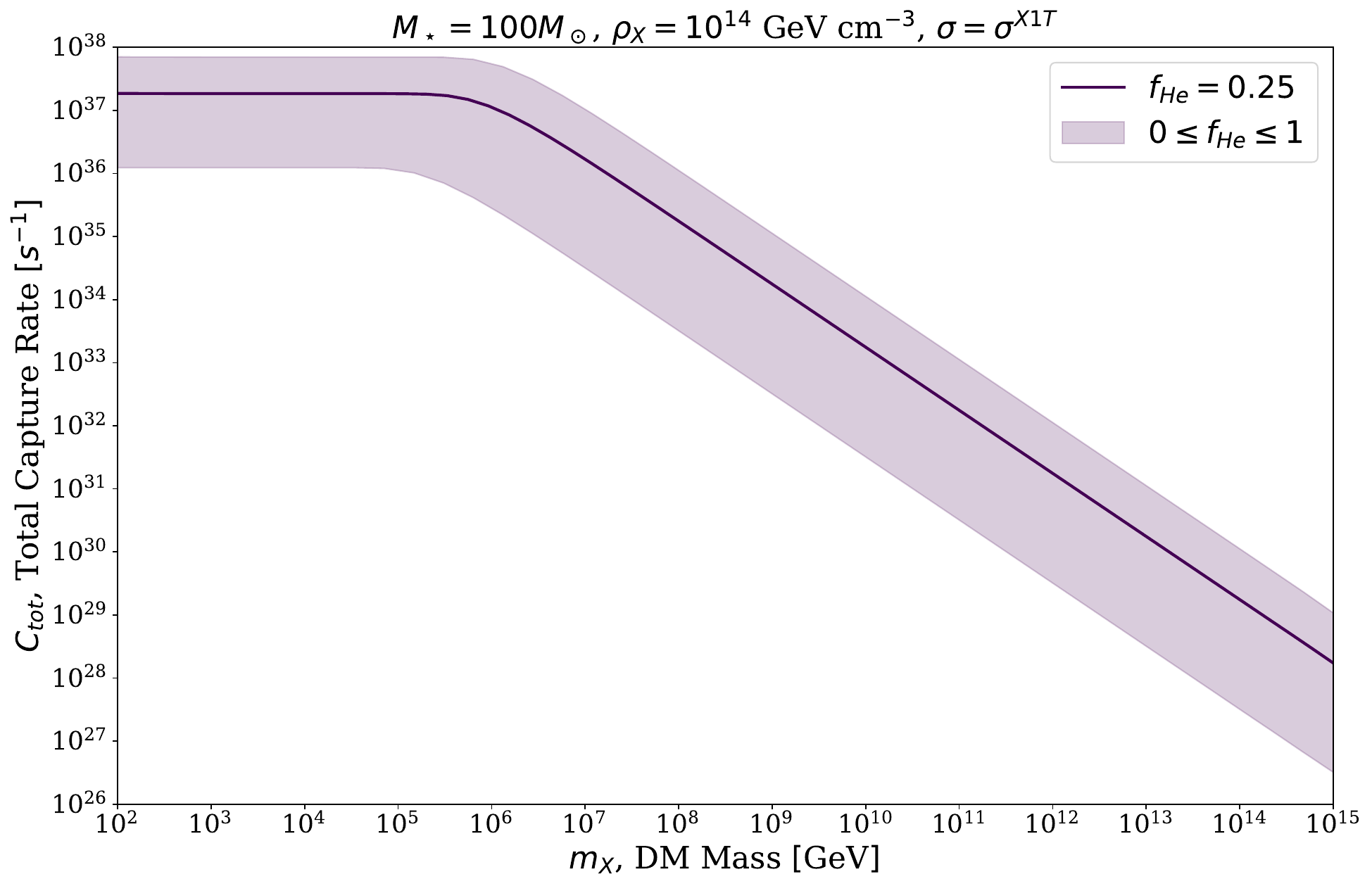}
\caption{Numerical verification of the two-component multiscatter formalism, demonstrating a capture rate in the appropriate range for a helium mass fraction of $f_{He} = 0.25$. The solid line shows the $f_{He} = 0.25$ case, within the appropriate range, along with a shaded region showing the possible capture rates for different mass fractions in the range $0 \lesssim f_{He} \lesssim 1$ using the two-component multiscatter formalism. The lower (upper) boundary of the shaded region represents the limiting case of $f_{He} = 0$ ($f_{He} = 1$).}
\label{fig:VerifyNumericallyCtot}
\end{figure}

We now present a consistency check to numerically verify that the multi-component multiscatter formalism produces capture rates in the range between the two extremes: $f_{He} = 0$ and $f_{He} = 1$. We adopt this range of fractions simply to verify that the formalism produces capture rates in the expected range and reduces to the single-component case in the correct limits. The true expected helium fraction at ZAMS is $f_{He} = 0.25$. In Sec. \ref{sec:TwoCompformalism}, we showed analytically how the multi-component formalism reduces to the single-component formalism in the limit of a pure object. Here, we show numerically that capture rates for two-component objects fall between the two limiting cases. To do this, we use the single-component formalism given by Eq.~(\ref{equation:CnSCMS}) to calculate the total capture rates for Pop~III parameters assuming non-physical cases of $f_{He} = 0$ and $f_{He} = 1$ along with mass fractions in the intermediary ranges using the two-component formalism. Our results can be seen in Fig.~\ref{fig:VerifyNumericallyCtot}, which shows the DM capture rate for a $100 M_\odot$ Pop~III star assuming various compositions. It is clear that the capture rate using the physically justified mass fraction $f_{He} = 0.25$, falls within the expected range given by the two limiting regimes. We thus demonstrate the validity of the multi-component multiscatter formalism, when applied to two-component objects. Moreover, note that, when compared to the case of $f_{He}=0$ (i.e. H alone), the capture rate in a realistic Pop~III star ($f_{He}=0.25$) is enhanced by roughly one order of magnitude.

\begin{figure}   [!thb]
\centering
\includegraphics[angle=0,scale=0.4]{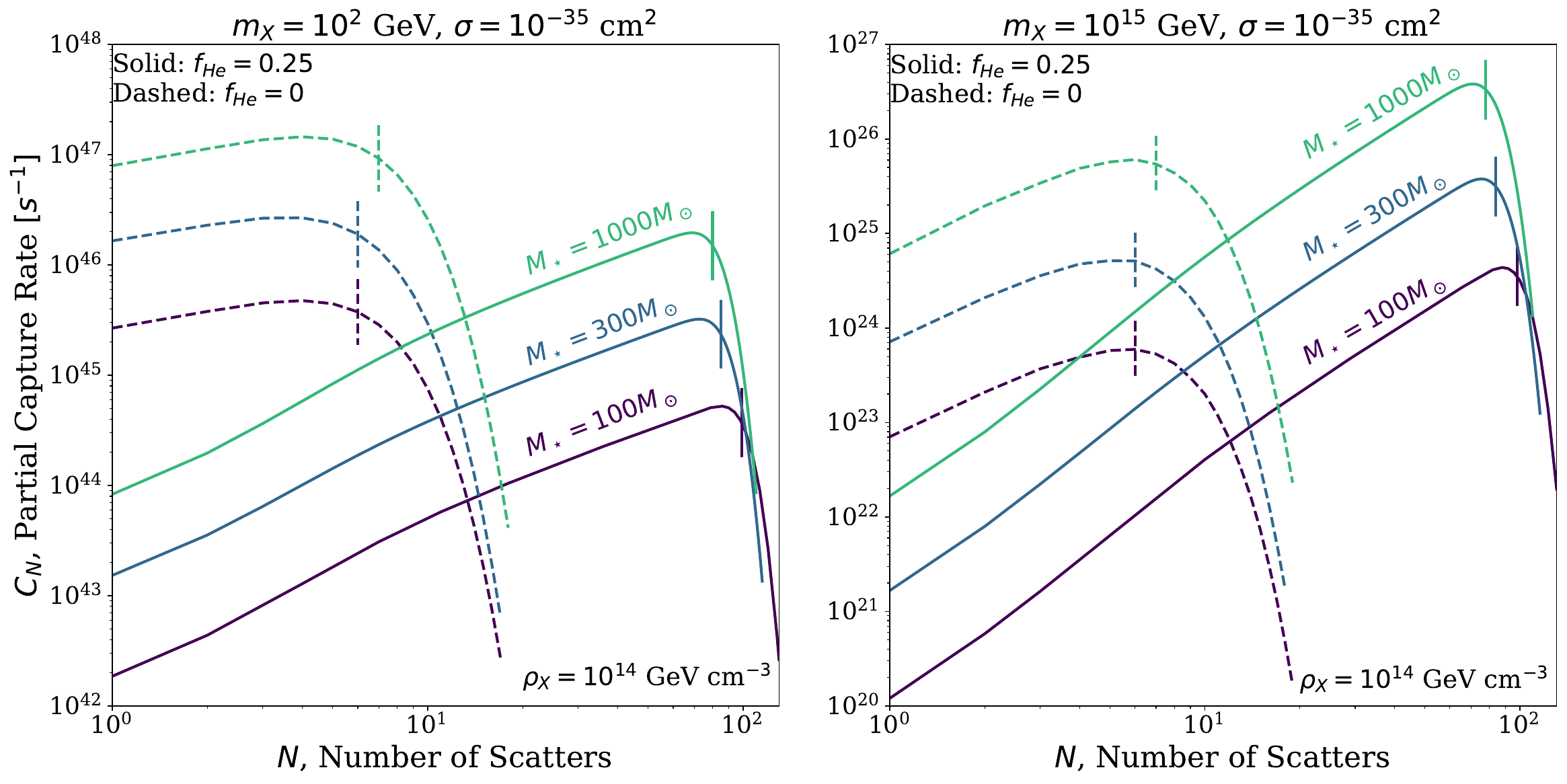}
\caption{Partial capture rate, $C_{N}$, plotted against the number of scatters for a $m_X = 10^{2}$ GeV (left panel) and and $m_X = 10^{15}$ GeV DM particle. In this plot, the solid lines represent the $f_{He} = 0.25$ case while the dashed lines represent $f_{He}=0$. The solid (dashed) vertical lines across the curves represent $\tauhe$ ($\tauh$) for $f_{He} = 0.25$ ($f_{He} = 0$). We remind the reader here that the optical depths, $\tauh$ and $\tauhe$, depend only on the cross section and the capturing object and are therefore invariant under changes in DM mass. This explains the same $N$ position of the vertical lines in both panels. The abrupt drop in the partial capture rate results from low probabilities of DM particles scattering $N > \tau$ times and demonstrates how the series defined in Eq. \ref{equation:CtotBasic} is truncated.}
\label{fig:CnN}
\end{figure}

In Fig.~\ref{fig:CnN}, we plot the partial capture rate $C_{N}$ from Eq.~(\ref{equation:CnSolved}), against the number of scatters, $N$, for pure-hydrogen and $\sim 25\%$ helium stars to demonstrate the effect helium has on the capture process. In both cases, we find that for large values of N, the partial capture rate begins to fall rapidly until the cutoff condition stipulated in Eqs.~(\ref{eq:Cutoff1}~-~\ref{eq:Cutoff2}) is reached. In the single-component case, this is because the probability of a DM particle scattering $N > \tau$ times is very low and so once the average number of scatters is surpassed, the capture rate is suppressed accordingly. In the multi-component case, naturally, it is the sum of optical depths, $\sum_i \tau_i$, that governs the cutoff condition. For Pop~III stars, helium dominates the capture process and thus the cutoff in Fig.~\ref{fig:CnN} for the $f_{He} = 0.25$ case occurs soon after $\tau_{He}$. In the case of Pop~III stars, we obtain the following scaling relationships for $\tauh$ and $\tauhe$:
\begin{equation}
  \tau_{H} = 10^{-5} \left(\frac{\sigma_{H}}{1.26 \times 10^{-40}~\text{cm}^2}\right) \left(\frac{M_{\star}}{M_{\odot}}\right) \left(\frac{R_{\odot}}{R_{\star}}\right)^2 \left(\frac{f_{H}}{0.75}\right),
\label{equation:tauScale1}
\end{equation}
\begin{equation}
    \tau_{He} = 10^{-3} \left(\frac{\sigma_{H}}{1.26 \times 10^{-40}~\text{cm}^2}\right) \left(\frac{M_{\star}}{M_{\odot}}\right) \left(\frac{R_{\odot}}{R_{\star}}\right)^2 \left(\frac{f_{He}}{0.25}\right) \left(\frac{\langle F^2(E_R)\rangle}{0.99}\right).
\label{equation:tauScale2}
\end{equation}
In most cases ($f_{He} \gtrsim 0.02$), $\tauhe > \tauh$ and the cutoff associated with capture from these stars happens soon after $N > \tauhe$ and long after $N > \tauh$. This can be seen in Fig.~(\ref{fig:CnN}), where the vertical lines represent the different values of $\tau$ for the different stars. In the multi-component case, it is $\tauhe$ that determines when the partial capture rates begin to fall off. We remind the reader that $f_H = 1$ Pop~III stars are not realistic, however we compare to the pure-hydrogen case to demonstrate clearly the effect of considering helium scatters on the capture rate, and to contrast with previous results that assumed hydrogen-only stars. It is important to note that these optical depths do not depend on DM mass, but only on the scattering cross section and capturing object.~Thus, changing the DM mass has no effect on the average number of times DM will scatter with each component and the associated probability of DM undergoing a specific combination of scatters (governed by $p_i(\tau_A)p_j(\tau_B)$). Rather, its effect is relevant for the probability of being captured after a specific scattering combination, which is encoded in the probability function $g_{ij}$ defined in Eq.~(\ref{eq:gij_integral}). Ultimately, increasing the mass of DM imposes a tighter cutoff on the velocity integral of Eq.~(\ref{equation:CnMulti}), which means less of the ambient DM is available for capture and the capture rate is lower, a fact expressed in Fig.~\ref{fig:CnN}.

\begin{figure}   [bht]

\centering

\includegraphics[angle=0,scale=0.45]{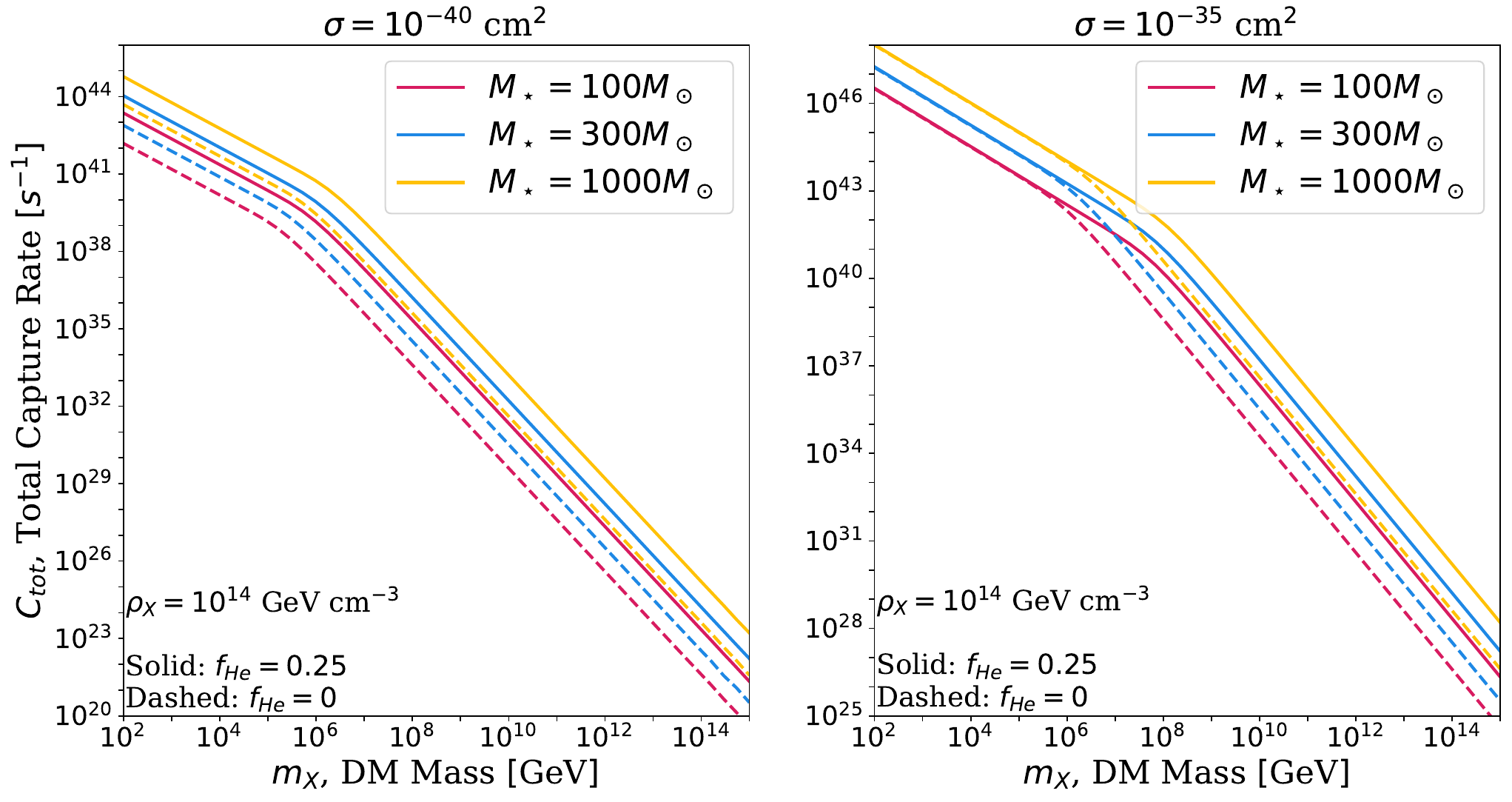}

\caption{Total numerical DM Capture rates for various Pop. III stars containing $\sim 25\%$ and $\sim 0\%$ Helium in a single-scatter (left panel) and a multi-scatter (right panel) scenario. For the left panel, the optical depths are in the ranges: $5.18\times 10^{-5} \leq \tauh \leq 5.32\times 10^{-5}$ and $7.9\times 10^{-4}\leq \tauhe\leq 9.8\times 10^{-4}$, and the right panel are the same up to a factor of $10^5$ since $\tauh,\tauhe\sim\sigma$ and $10^{-35} = 10^5\times 10^{-40}$. The dashed lines represent the pure-hydrogen stars while the solid lines are those containing $\sim 25\%$ helium. There is an enhancement of the total capture rate across the entire DM mass range examined for single scattering and for the high-mass region of the multiscattering case. This is due to a greater average number of scatters and greater average energy lost per scatter in these regions. The low-mass multiscattering regime produces identical capture rates to pure-hydrogen stars as the low DM mass and high number of scatters guarantees all transiting particles become gravitationally bound, irrespective of the star's total helium content.}
 
\label{fig:CtotMchiNum25vs0}
\end{figure}

An intriguing feature of Fig.~\ref{fig:CnN} is the contrast in helium's effect on the partial capture rate for WIMP versus superheavy dark matter. For the $m_X = 10^{15}$ GeV DM particle, the peak in the partial capture rate for a $f_{He} = 0.25$ star surpasses the peak of the $f_{He} = 0$ object by about an order of magnitude. Overall, this leads to an enhancement of the total capture rate in the high-mass, multiscatter range, explicitly shown in the high mass region of the right panel in Fig.~\ref{fig:CtotMchiNum25vs0}. In contrast, the peak is lower for the $f_{He} = 0.25$ case when considering $m_X = 10^2$ GeV DM. One might then naively conclude that the total capture rate would be higher for the $f_{He}= 0$ case. In actuality, low-mass multiscattering of hydrogen and helium produces identical total capture rates to pure-hydrogen scattering, as all transiting dark matter particles become captured, a fact demonstrated in the low mass region of the right panel in Fig.~\ref{fig:CtotMchiNum25vs0}. This is not obvious in Fig.~\ref{fig:CnN}, but it is the case that the lower peak in the partial capture rate for $f_{He} = 0.25$ in the left panel is compensated by the higher cutoff of $N$, leaving the capture rate identical. To show this is the case analytically, we can take the limit of $\tau_H, \tau_{He} \to \infty$ in Eq.~(\ref{eq:Ctot_TauBig_RvBig}), an analytic form of the total capture rate that is derived in Appendix~\ref{section:AnalyticExpressions} for $\tau_H, \tau_{He} \gg 1$ and $\langle \frac{3(v_{ij}^2 - v_{esc}^2)}{2 \vbar^2}\rangle \gg 1$.\footnote{This is the negative of the exponent found in the partial capture rate given in Eq.~(\ref{equation:CnSolved}), and is defined as $R_v$ in Appendix~\ref{section:AnalyticExpressions}. Recall that $v_{ij}$ is the DM velocity after $i$ hydrogen scatters and $j$ helium scatters, and thus this parameter depends on the kinematics of the collision and the masses involved. For low DM mass undergoing multiple scatterings, it is shown in Fig.~\ref{fig:Ctot_NumvsAnalytic_BigTau} that the average of $R_v$ is very large compared to unity, and thus the exponential term is suppressed accordingly.} Doing so gives the following expression: 
\be
C_{tot} \approx \pi R_\star^2\left[3 \sqrt{\frac{6}{\pi}} n_X\left(\frac{2}{9}\vbar + \frac{v_{esc}^2}{3\vbar}\right)\right] = \pi R_\star^2 \times \int_{v_{esc}}^{\infty} dw \frac{f(u)}{u^2} w^3,
\label{eq:Ctot_tauBigLimit}
\ee
which one can verify is  the incident flux of DM on the star for a Maxwell-boltzmann distribution of the form: $f(u) du = 3 \sqrt{6/\pi} \frac{n_X u^2}{\vbar^3} \text{Exp}(-\frac{3 u^2}{2 \vbar^2}) du$. Notice that Eq.~(\ref{eq:Ctot_tauBigLimit}) contains only information on the star's mass and radius, irrespective of its contents and thus represents a geometric capture rate of DM. It is thus natural to expect the total capture rates for varying helium content to be identical in this regime as the high number of scatters and kinematics of the collisions guarantee that all transiting particles become gravitationally bound.

We will now show the effects of including $\sim 25\%$ helium in Pop~III stars on the total capture rates by comparing our results to a pure-hydrogen case in the single scattering and multiscattering regimes. Fig.~\ref{fig:CtotMchiNum25vs0} shows a direct comparison of the total capture rate when considering pure-hydrogen versus $\sim 25\%$ helium Pop~III stars in the optically thin ($\sigma = 10^{-40}~\text{cm}^2$ and $\tauh, \tauhe \ll 1$) and the optically thick limits ($\sigma = 10^{-35}~\text{cm}^2$ and $\tauh, \tauhe \gg 1$). Across most of the parameter space, the inclusion of helium leads to an enhancement of the total capture rate by about an order of magnitude. The exception is low-mass multiscattering which, as demonstrated in Eq.~(\ref{eq:Ctot_tauBigLimit}), leads to a total capture rate that does not depend on the helium content of the star. This is merely a geometric capture rate of DM, where the scattering cross section is high enough and the DM mass low enough that all transiting particles become trapped. We note here that while it is informative to study capture in the low-mass multiscatter regime, as this may be relevant for other objects, in the present study of Pop~III stellar capture of DM masses $m_X \gtrsim 10^2$ GeV, it represents an unrealistic scenario as the entire range in which it occurs ($10^2~\text{GeV}\lesssim m_X \lesssim 10^7~\text{GeV}$ and $\sigma \gtrsim 10^{-35}~\text{cm}^2$) is currently ruled out by direct detection experiments (Fig.~\ref{fig:sigma-mx_bounds} shows this region as excluded by the XENON1T experiment).

In all other cases (high-mass multiscattering, low-mass single-scattering, and high-mass single-scattering), the effect of helium is to enhance the total capture rate. To see why this happens, consider the enhancement of the DM-nucleon cross section when considering interactions with helium nuclei as shown in Eq.~(\ref{eq:sigmaHe}). This leads to a higher average number of scatters for DM particles incident on the star as demonstrated in Eq.~(\ref{equation:tauScale2}), making capture more likely. While there is a suppression of the cross section due to the form factor, this effect is negligible at the energies considered here. Furthermore, because helium nuclei are more massive than hydrogen nuclei, collisions with helium will, on average, cause DM particles to lose a larger fraction of their energy and thus lead to a greater likelihood of capture after a given number of scatters. As an example, consider the average relative energy lost by a DM particle in a single collision with helium versus hydrogen:
\be
    \epsilon = \frac{\langle z_{He}\rangle \beta_+^{He}}{\langle z_{H}\rangle \beta_+^{H}} \approx 4.
\ee
These factors combined cause the capture rate to be enhanced for the single scatter regime for all masses and the high-mass multiscatter regime.

We now go on to show how to calculate the DM luminosity in Pop~III stars assuming a self-annihilating model of DM. The following differential equation governs the total number of DM particles in the star as a function of time:
\be
    \dot{N} = C - \Gamma_A,
\ee
 where $C$ is the DM capture rate and $\Gamma_A$ is the annihilation rate of DM in the star. Note that we have neglected the effects of DM evaporation due to its sub-dominance in Pop~III stars for DM masses above the evaporation mass $m_{evap} \sim 10^{-2}$ GeV (See \cite{Ilie:2020PopIIIa} for estimates of the evaporation mass in Pop~III stars as well as detailed numerical and analytic calculations of the evaporation rates of DM off of Pop~III stars in the sub-GeV regime). For WIMPS, Ref.~\cite{Freese:2008cap} shows that an equilibrium between capture and annihilation ($\dot{N} = 0$) is reached quickly on timescales  much less than the star's age and Ref.~\cite{Ilie:2019} shows this is the case for higher DM masses assuming a distribution where most of the DM particles are captured near the core of the star. Equilibrium is also achieved for other DM models in the sub-GeV region, as discussed in \cite{Ilie:2020PopIII}. When equilibrium is reached, the annihilation rate equals the capture rate and DM can provide a stable source of luminosity modelled by:
\be\label{eq:Ldm_equalCmx}
    L_{DM} = f \Gamma_A m_X = f C m_X ,
\ee
where $f$ is the fraction of annihilation products thermalized in the star, not to be confused with the H or He fractions. Following Ref.~\cite{Ilie:2020PopIII}, we assume $f = 1$. Given the enhancement in DM capture through the modelling of helium scatters, there is a corresponding increase in the DM luminosity as $L_{DM} \sim C$. As we shall see in the next section, this leads to tighter upper bounds on Pop~III stellar masses for a given set of DM parameters.

\section{Constraints on Pop~III stellar mass from Dark Matter capture}\label{sec:MaxMstar}
The additional luminosity provided by DM annihilation leads to the possibility of constraining the mass of Pop~III stars. Stars which have become radiation pressure dominated will have their luminosity scale linearly with the mass of the star in the Eddington limit. This means that for a given stellar mass $M_\star$, the Eddington luminosity cannot be exceeded as further mass accretion is prevented by the outward radiation pressure. We can define the Eddington luminosity as a function of a star's mass and atmospheric opacity:
\be
    L_{Edd} = \frac{4 \pi c G M_\star}{\kappa_\rho},
\ee
where $c$ is the speed of light, $G$ is the gravitational constant, $M_\star$ is the stellar mass, and $\kappa_\rho$ is the opacity of the stellar atmosphere. We point out that the first stars have low metallicity and hot atmospheres, and therefore the opacity arises from Thompson electron scattering. The atmospheric opacity from Thompson scattering depends only on the fraction of hydrogen in the star $f_{H}$, by the following relationship: $\kappa = 0.2(1 + f_{H})$ cm$^{2}$ s$^{-1}$. Taking $f_{H} = 0.75$ for ZAMS Pop~III stars gives the following scaling relationship for the Eddington luminosity:
\be\label{eq:L_eddington}
    L_{Edd} = 3.71 \times 10^{4} (M_\star/M\odot) L_\odot.
\ee
We place projected upper bounds on the mass of Pop~III stars by requiring that observed Pop~III stars will respect the Eddington limit:
\be
    L_{Edd}(M_{max}) \leq L_{nuc}(M_{max}) + L_{DM}(M_{max}),
    \label{eq:MaxMassStrong}
\ee
where $L_{nuc}$ is the star's nuclear luminosity (See Table \ref{table:StellarData}). To model the nuclear luminosity for any Pop~III stellar mass between $\sim 1- 1000\Msun$, we use an interpolating fit for the stellar mass and luminosity in Table \ref{table:StellarData} given by \cite{Ilie:2020PopIIIa}:
\be
    L_{nuc} \simeq 10^{\frac{\log(3.71 \times 10^4 L_\odot s/erg)}{1 + exp(-0.85\log(x) - 1.95)}} \times x^{\frac{2.01}{x^{0.48} + 1}} erg/s,
    \label{equation:LnucFit}
\ee
where $x \equiv \frac{M_\star}{M_\odot}$. It is important to note that in placing bounds on stellar masses in this way, we assume that the additional DM luminosity does not influence significantly the structure of the star by way of the mass-radius homology relations. This is a reasonable approximation, at least for the higher end of the stellar mass range considered, since those stars are already almost Eddington limited, so it will take only a nudge from DM heating to have them reach the luminosity limit. However, we plan to test this in the future by incorporating the effects of DM heating into the stellar evolution code \texttt{MESA}.

\begin{figure} [bht]

\centering

\includegraphics[angle=0,scale=0.45]{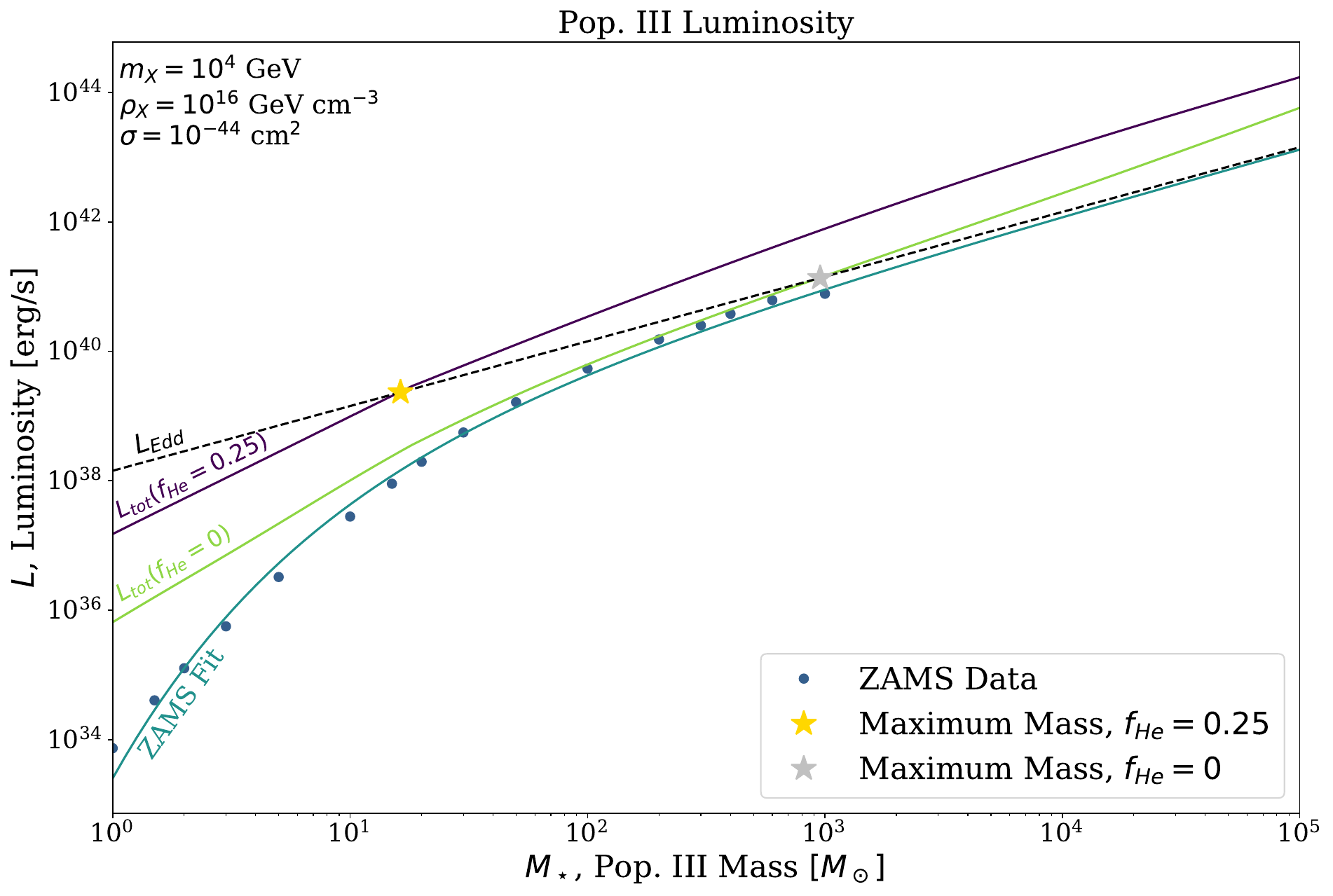}

\caption{Luminosity against stellar mass, demonstrating the effect DM luminosity has on maximum Pop~III mass by virtue of the Eddington limit. The circular points represent luminosities of zero age main sequence Pop~III stars from simulations with the solid blue line being a fit of this data given by Eq.~(\ref{equation:LnucFit}). The dashed black line represents the Eddington luminosity as a function of stellar mass as given by Eq.~(\ref{eq:L_eddington}). The purple line represents the total luminosity including effects of DM annihilation where DM particles scatter off hydrogen and helium. The green line is the total luminosity when considering hydrogen scatters only. Notice the significant increase in the total luminosity of the star as more DM particles become captured and annihilate when considering helium scattering. The stars represent the mass at which the Eddington limit is saturated and are thus projected upper bounds on Pop~III mass for the given DM parameters.}
 
\label{fig:LvsMstars}
\end{figure}

In Fig.~\ref{fig:LvsMstars}, we plot stellar luminosity against stellar mass for a wide range of Pop~III stars. For nuclear luminosity, the data from Table \ref{table:StellarData} is presented along with the fitting formula for other masses given by Eq.~(\ref{equation:LnucFit}). DM luminosity is calculated from Eq.~(\ref{eq:Ldm_equalCmx}), where the radius of the Pop~III star of a given mass is inferred from the following homology relations derived in \cite{Ilie:2019}:
\be\label{eq:HomoRels}
  \frac{\Rstar}{R_\odot}\approx
  \begin{cases}
                                   0.88\left(\frac{M_\star}{\Msun}\right)^{0.20} & \text{if $M_\star \lesssim 20M_\odot$} \\
                                   0.32\left(\frac{M_\star}{\Msun}\right)^{0.55} & \text{if $M_\star \gtrsim 20M_\odot$.} \\
  \end{cases}
\ee
These relations were inferred by fitting the data in Table \ref{table:StellarData} in two distinct regimes separated by the natural breaking point at $M_\star \sim 20 M_\odot$, shown explicitly in Fig.~2 of Ref.~\cite{Ilie:2019}. For this reason, a broken power law behavior for the total luminosity is observed in Fig.~\ref{fig:LvsMstars}.
We have also included in Fig.~\ref{fig:LvsMstars} the maximum mass resulting from nuclear and DM luminosity summing to the Eddington limit, i.e., the stellar mass solving Eq.~(\ref{eq:MaxMassStrong}). Firstly, notice the effect of considering helium scatters on the total luminosity. The introduction of helium causes the luminosity to increase across all DM mass ranges and so for the case of $m_X = 10^{4}$ GeV DM in Fig.~\ref{fig:LvsMstars}, the total luminosity curve when $f_{He} = 0.25$ is shifted upwards for all stellar masses. This means there is a stronger heating effect from DM for all stellar masses when modelling the presence of helium in their composition. This is justified by the enhancement of DM capture when helium scatters are considered and the relationship $L_{DM} \sim C_{tot}$ arising when capture and annihilation are in equilibrium. The effect of helium can also be seen through more stringent bounds on Pop~III stellar masses. For a given stellar mass, the additional DM heating from the presence of helium pushes the star closer to the Eddington limit than if one neglects helium scatters. This leads to tighter bounds on Pop~III stellar masses. In Fig.~\ref{fig:LvsMstars}, for the sake of easy visibility, we have plotted the maximum mass from DM and nuclear luminosity for $f_{He} = 0$ and $f_{He} = 0.25$ stars when considering $m_X = 10^{4}$ GeV DM. Notice the downward shift in the maximum mass, as expected from the enhanced DM luminosity.

\begin{figure}[bht]
    \centering
    \includegraphics[angle = 0, scale = 0.45]{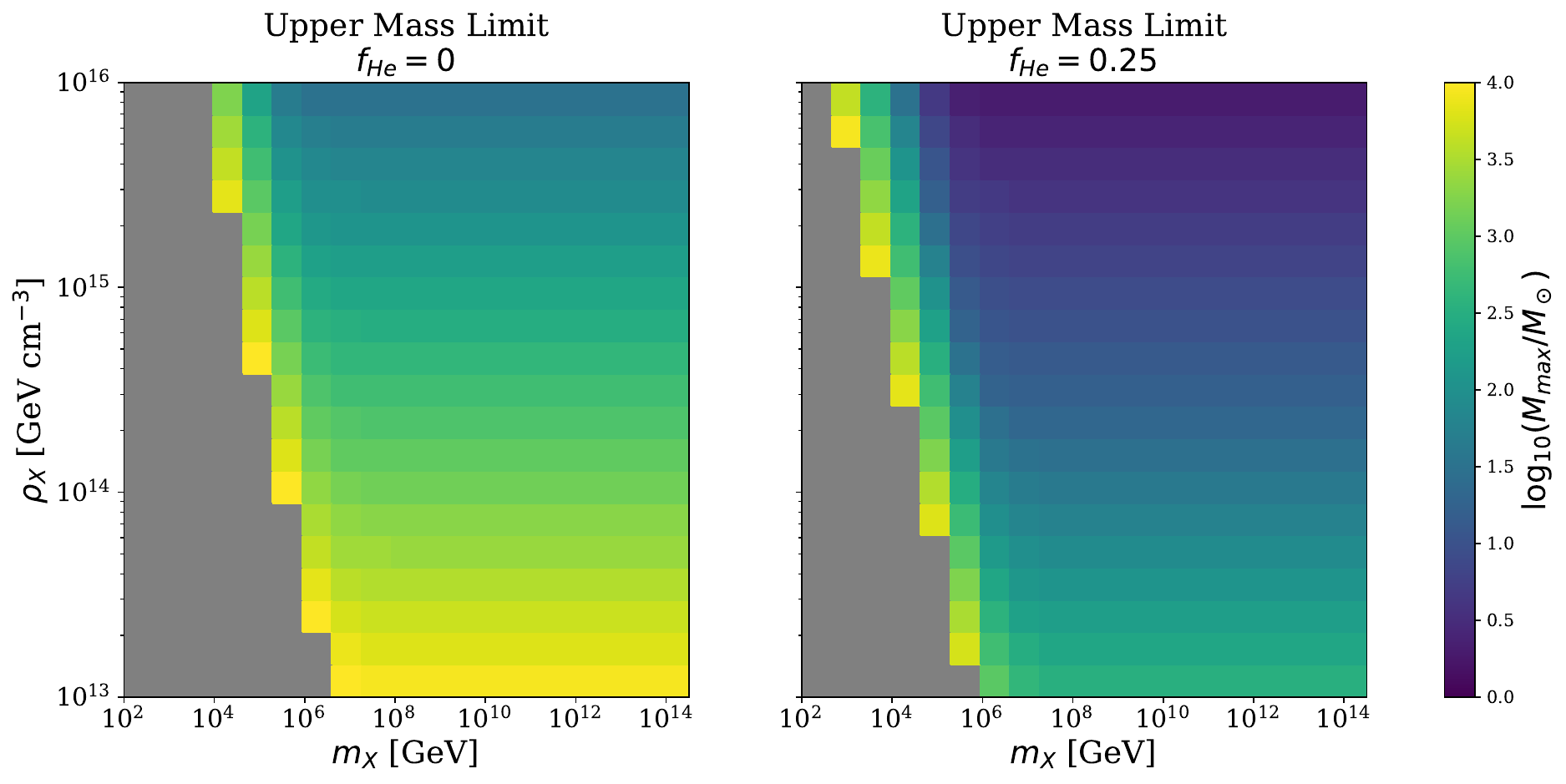}
    \caption{Maximum Pop~III stellar mass as a function of ambient DM density and DM mass by virtue of the Eddington limit. We present projected upper bounds on Pop~III stellar masses for pure-hydrogen and $\sim 25\%$ helium cases to demonstrate the tightening of bounds from modelling helium capture. The color at each point represents the maximum Pop~III mass for those given parameters and the gray region represents upper bounds $ \gtrsim 10^4 M_\odot$, where other processes, such as radiative feedback, will likely limit stellar mass \cite{Stacy:2016}. For both cases, we take $\sigma_0^{SI-p}$ from XENON1T bounds~\cite{Aprile:2018} as shown in Eq.~(\ref{Eq:X1Tbounds}). The higher DM luminosities resulting from modelling helium capture leads to tighter bounds on Pop~III masses for all given parameters, and even allows for constraining power in previously unavailable regions of DM mass-density parameter space.}
    \label{fig:MaxMass_Strong}
\end{figure}

To visualize the effects of varying DM parameters, namely DM mass and density, on the maximum Pop~III stellar mass, we plot the maximum stellar mass in the DM density-mass parameter space for pure-hydrogen and $\sim 25 \%$ helium stars in Fig.~\ref{fig:MaxMass_Strong}. Note that we have excluded bounds for stellar masses exceeding $\sim 10^4$ $M_\odot$ via the gray region on the plot as radiative feedback and fragmentation of the gas cloud would likely prevent Pop~III stars from reaching these masses. This allows us to see which areas of the parameter space are useful in reasonably constraining Pop~III stellar mass and what these bounds are for a set of parameters. In both cases, we can see the effects of varying DM mass on the max Pop~III mass. Firstly, note that there exist two distinct regimes, as expected from the behavior of DM capture against DM mass seen in Fig.~\ref{fig:VerifyNumericallyCtot} and expressed analytically in Eq.~(\ref{equation:CtotApprox_Pop3}), where $\sigma$ is taken from XENON1T bounds~\cite{Aprile:2018} for $m_X \gtrsim 10^2$ GeV in Eq.~(\ref{Eq:X1Tbounds}). For lower DM masses, increasing the mass of DM leads to greater DM luminosity and so tighter forecasted bounds on stellar masses as evidenced by the darkening in the lower DM mass region in Fig.~\ref{fig:MaxMass_Strong}. For higher DM masses, the luminosity is independent of DM mass and so the color becomes constant past a certain DM mass. With regards to DM density, we can see that an increase in the density leads to tighter bounds on stellar masses across the entire range. This can be explained by the fact that the capture rate scales with density across the entire DM mass range and thus higher densities lead to greater DM capture and thus luminosity. This greater luminosity pushes a star to the Eddington limit at lower stellar masses.

The effect of helium on the ability to constrain Pop~III stellar parameters is significant. We have demonstrated that tighter bounds can be placed across the entire parameter space and even allows for us to constrain parts of parameter space that were previously unavailable. This result is expected due to the enhancement of DM capture and luminosity by accounting for helium in the star.

\section{Bounds on Dark Matter Properties from Pop~III stars}\label{sec:BoundsOnDM}
In this section, we demonstrate a method for constraining DM properties through the observation of Pop~III stars and discuss the effect that their helium content has on the projected bounds. Previous work has been done on placing projected bounds on DM properties utilizing this method under the assumption of pure-hydrogen stars \cite{Freese:2008cap,Ilie:2020PopIIIa,Ilie:2020BNFa}. For that reason, we will compare our results to those obtained previously to demonstrate the effect of helium on the constraining power of the method. The main idea for placing constraints on DM properties is the following: if we are to observe Pop~III stars, implying they obey the Eddington limit, what can we learn about DM properties? To answer this question, we point to the following equation:
\be
    C_{tot} \leq \frac{L_{Edd} - L_{Nuc}}{f m_X},
    \label{eq:Bounds}
\ee
which is simply Eq.~(\ref{eq:MaxMassStrong}) rearranged to demonstrate the process we use to take bounds. Firstly, note that this equation stems from an inequality of the Eddington luminosity with the star's total luminosity. Previously, we demonstrated that projected bounds can be placed on Pop~III masses using this inequality for a given set of DM parameters. We are now re-framing this equality to pose the following question: if we observe Pop~III stars, what can we learn about DM parameters? This question provides the basis of the method we intend to use to place projected bounds on DM properties for a set of Pop~III parameters. 

Most generally, this method constrains the combination of two parameters $\rho_X \times \sigma$. Constraining either parameter independently requires making assumptions on the other. For the constraints on $\sigma$, a range of DM densities representative of adiabatically contracted NFW profiles are used, such as those found in \cite{Freese:2008ds, Freese:2008dmdens, Ilie:2020PopIIIa, Ilie:2020BNFa}. The uncertainty in its value is represented by a band of constraints. For constraining $\rho_X$, the current best bounds on $\sigma$ given by the XENON1T experiment \cite{Aprile:2018} up to its maximum sensitivity at the neutrino floor~\cite{Billard:2013} is used. For the case of $\rho_X \times \sigma$ projected bounds, no assumptions are made on either parameter and thus these are the most general bounds placed. As discussed above, to place these bounds we require stellar parameters. Pop~III stars have yet to be confirmed via observation and so parameters given by simulations outlined in Table \ref{table:StellarData} are used. We also utilize parameters given by \cite{Vanzella:2020}, where a potential Pop~III stellar complex is observed at $z \sim 6.629$. They utilize the same simulated Pop~III masses as those in this paper to approximate the number of stars found in this complex. For more information on the possibilities of Pop~III formation at these redshifts, see Ref~\cite{Mebane:2018}.

Presented first are the most general projected bounds one can place on the combination of parameters: $\rho_X \times \sigma$. To do this, analytic expressions for the total capture rate are used. Our result can be found in Eq.~(\ref{equation:RhoSig}) with a detailed derivation in Appendix \ref{section:AnalyticExpressions}. 
\begin{figure}[!htb]
    \centering
    \includegraphics[angle = 0, scale = 0.45]{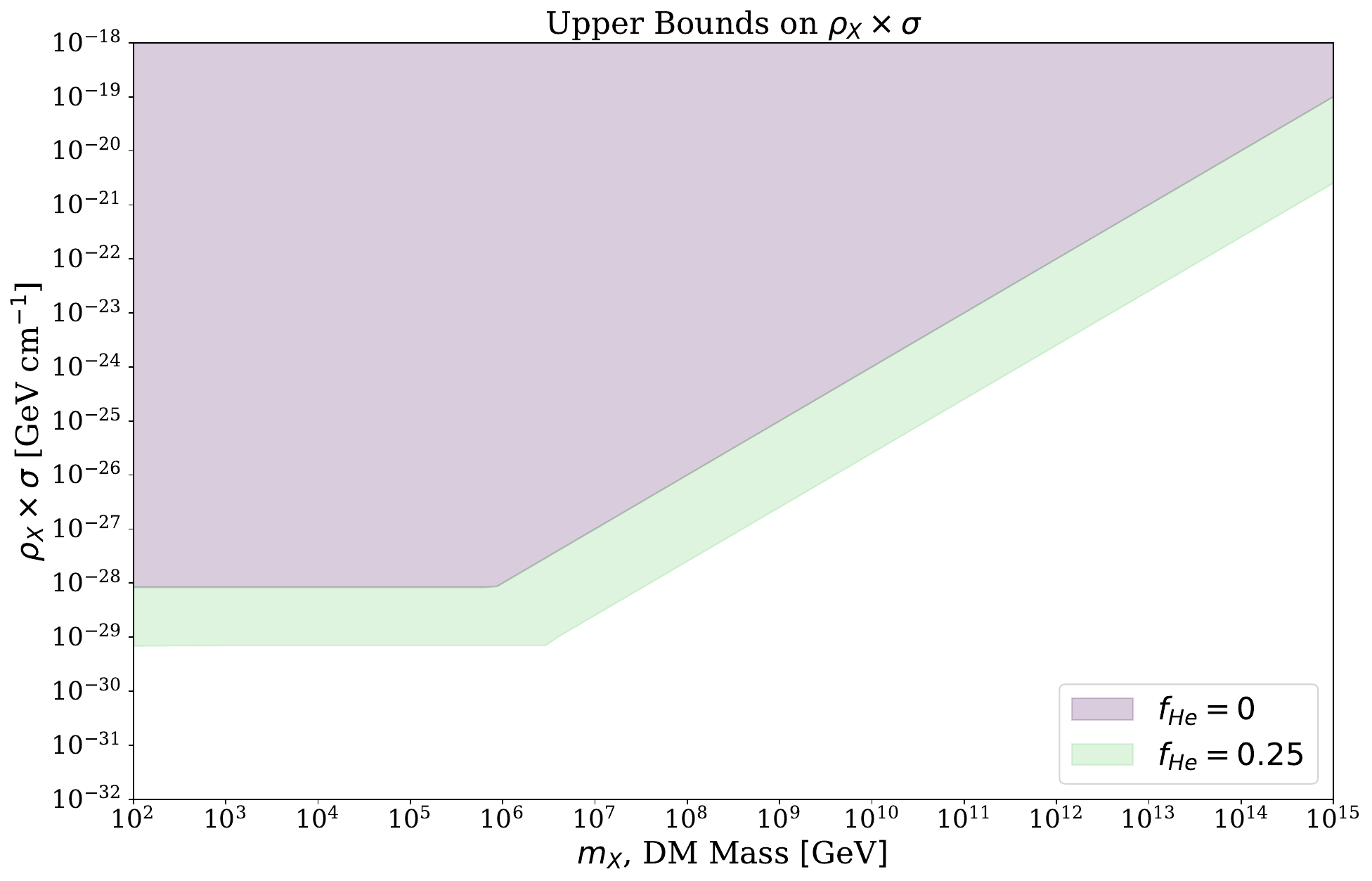}
    \caption{Projected upper bounds on the combination of ambient DM density and DM-proton scattering cross section, $\rho_X \sigma$, placed based on the potential observation of a $M_\star = 1000 M_\odot$ Pop~III star. The purple region represents the excluded region of this parameter space when considering scattering off of hydrogen nuclei only. The more stringent bounds, seen in green, are those placed when one considers the enhanced rate of DM capture due to scattering with helium nuclei. These bounds are the most general bounds we place based solely on the observation of Pop~III stars. Breaking this degeneracy requires an assumption on either $\rho_X$ or $\sigma$.}
    \label{fig:rhosigma-mx_bounds}
\end{figure}
In Fig.~\ref{fig:rhosigma-mx_bounds} we plot upper bounds on this combination of parameters from the potential observation of a $M_\star = 1000 M\odot$ Pop~III star. The shaded regions represent the excluded values of this product solely from observing Pop~III stars. As expected, modelling helium scatters leads to bounds that are more stringent due to the enhancement of DM capture. As discussed previously, since $L_{DM} \sim C_{tot}$ when equilibrium occurs, a larger DM capture rate, resulting from helium's high scattering cross section and mass relative to hydrogen, leads to a greater luminosity. Thus, modelling helium allows for more exclusion power as larger swaths of parameter space violate the Eddington limit.

\begin{figure}[htb]
    \centering
    \includegraphics[angle = 0, scale = 0.45]{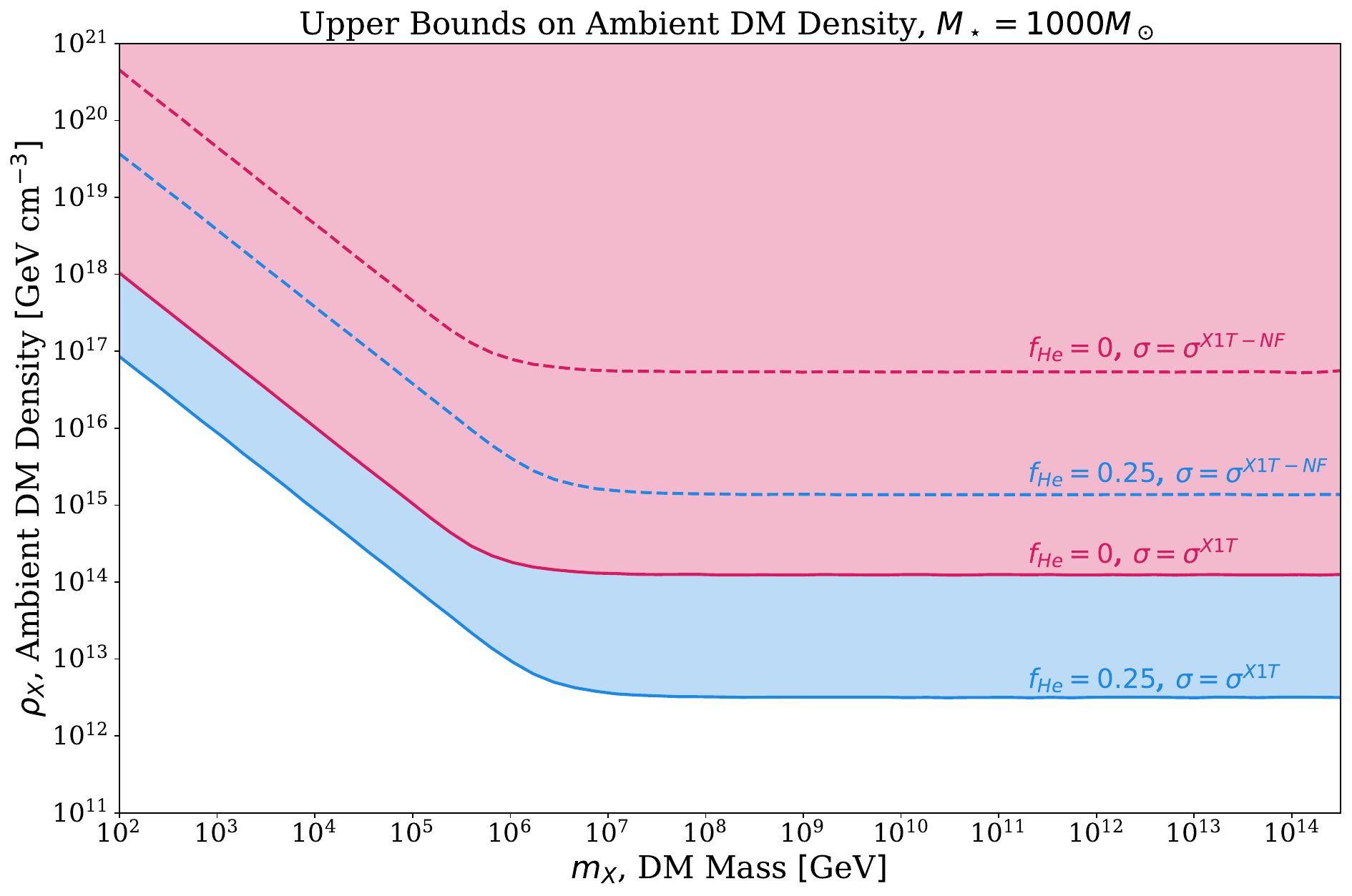}
    \caption{Projected upper bounds on DM density in the capturing region surrounding Pop~III stars obtained assuming DM-proton scattering cross sections in the region  below the current XENON1T sensitivity limit \cite{Aprile:2018} and above the neutrino floor~\cite{Billard:2013}. 
    Including helium (seen by the bounds in blue) leads to more stringent constraints on this key DM parameter relative to hydrogen-only scattering (seen by bounds in red). The solid lines represent upper bounds taking $\sigma$ at the current XENON1T limit while the dashed lines take $\sigma$ at the neutrino floor where the XENON1T experiment would reach maximum sensitivity~\cite{Billard:2013}}
    \label{fig:rho-mx_bounds}
\end{figure}

Next, the density of DM particles in the region surrounding Pop~III stars is constrained. This is a key parameter that is poorly constrained by observational evidence.~As done in this paper, it is standard to use the adiabatic contraction technique to estimate this density parameter by modelling the collapse of the baryonic core through the conservation of adiabatic invariants~\cite{Blumenthal:1985,Young:1980,Gendin:2011}. While there do exist numerical simulations of this process \cite{Abel:2001}, they are limited in resolution and cannot probe far enough inwards to the edge of the baryonic core.~We thus present a way to place upper bounds on this parameter through the observation of Pop~III stars and assuming knowledge of the proton-DM cross section from direct detection experiments. If DM is to be identified by direct detection, with current techniques, the cross section will be restricted to a relatively narrow swath of parameter space, squeezed between the current exclusion limits and the so called neutrino floor~\cite{Billard:2013}. So, for $\sigma$ we take possible values in the region described above. Our main results are seen in Fig.~\ref{fig:rho-mx_bounds} for the case of a $M_\star = 1000 M_\odot$ Pop~III star.

From Fig.~\ref{fig:rho-mx_bounds}, it is evident that modelling helium leads to an order of magnitude increase in constraining power across all DM masses. Again, this is due to higher DM capture rates associated with helium scattering and the order of magnitude increase in DM capture rates across all parameters. For the physically realistic case, $f_{He} = 0.25$, the observation of a $M_\star = 1000 M_\odot$ Pop~III star implies an upper limit on the ambient DM density as low as $\sim 10^{13}$ GeV cm$^{-3}$ for DM masses $\gtrsim 10^{7}$ GeV and up to $\sim 10^{17}$ GeV cm$^{-3}$ for $m_X \simeq 10^2$ GeV when using current XENON1T bounds on $\sigma$. If the XENON1T experiment reaches maximum sensitivity without a detection, these bounds become weaker due to the inverse relationship between $\rho_X$ and $\sigma$ in when considering DM luminosity. At maximum sensitivity, our projected bounds fall to $\sim 10^{15}$ GeV cm$^{-3}$ for DM masses $\gtrsim 10^{7}$ GeV and up to $\sim 10^{19}$ GeV cm$^{-3}$ for $m_X \simeq 10^2$ GeV. We emphasise again that the ``bounds'' presented in Fig.~\ref{fig:rho-mx_bounds} make the additional assumption that direct detection experiments would have detected DM anywhere within the allowed parameter space, and above the ``neutrino floor.'' Perhaps counter-intuitively, as the bounds on $\sigma$ become tighter, there is a loss in constraining power of our method on $\rho_X$, due to the inverse relationship between the two parameters in this context. This is supported by the dashed lines in Fig.~(\ref{fig:rho-mx_bounds}), where tighter bounds on $\sigma$ by the XENON1T experiment imply weaker bounds on $\rho_X$ using our constraining method.

\begin{figure}[thb]
    \centering
    \includegraphics[angle = 0, scale = 0.45]{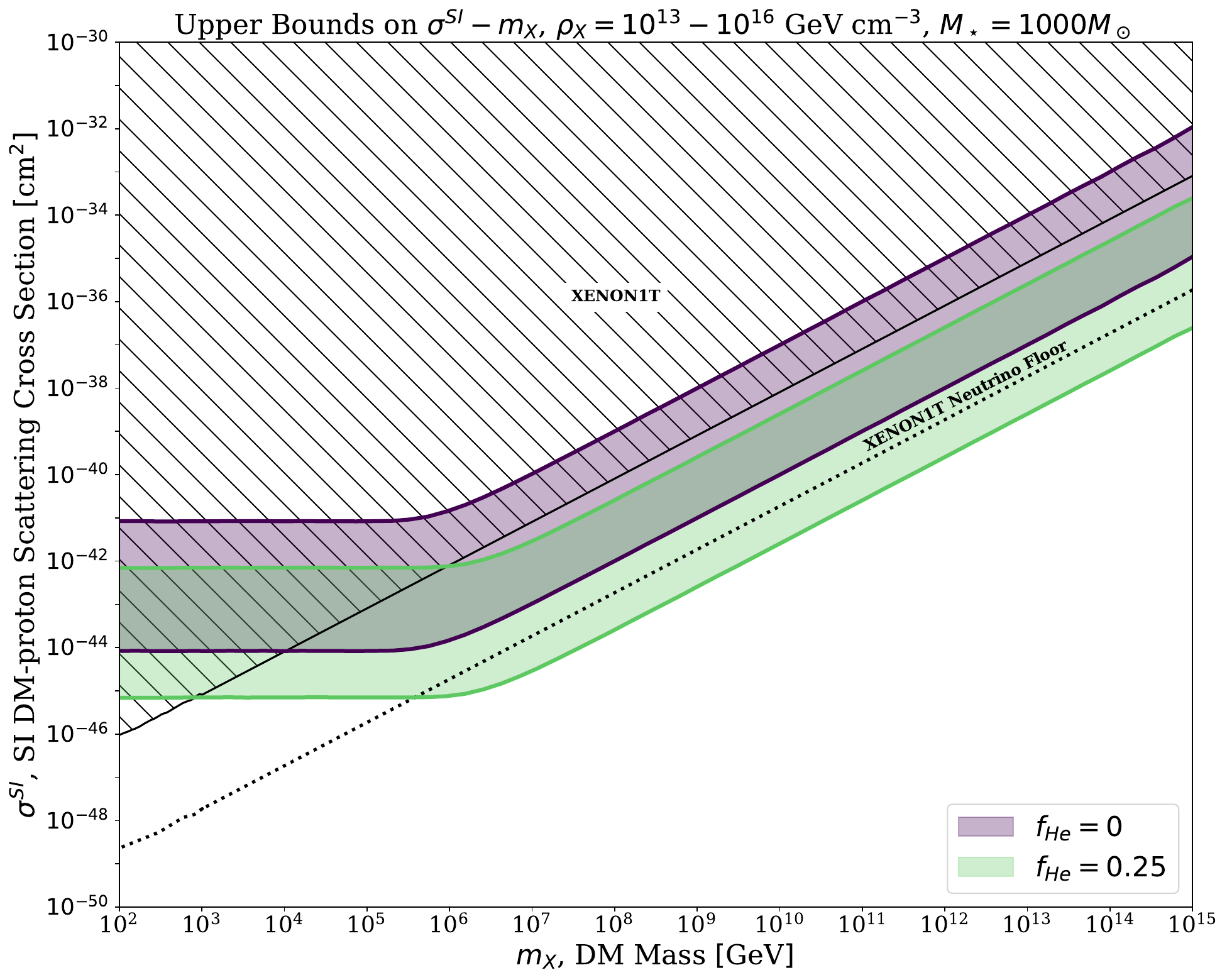}
    \caption{Projected upper bounds on the DM-proton scattering cross section from the observation of $M_\star = 1000 M_\odot$ Pop~III stars. We consider two cases for the Helium fraction, labeled in the legend. The bands for each case represents the uncertainty in the ambient DM density. We take this parameter to range from $10^{13} - 10^{16}$ GeV cm$^{-3}$. 
    The bounds placed when considering helium scatters are deeper by a factor of $\sim 10$ resulting from the enhanced capture rates due to helium scatters.}
    \label{fig:sigma-mx_bounds}
\end{figure}

We next present the most exciting application of our formalism: projected bounds on the DM-proton scattering cross section. To do so, we are using ambient DM densities ranging from $10^{13}$ to $10^{16}$ GeV cm$^{-3}$, as estimated using adiabatic contraction (See Appendix E in Ref.~\cite{Ilie:2020PopIIIa} for details on this).~As discussed previously, the modelling of helium scattering along with hydrogen scattering for capturing DM particles ultimately led to more stringent constraints on $\rho_X \times \sigma$. Thus, for a given DM density, we expect that upper bounds on $\sigma$ will become tighter for the models including helium scattering. This result is confirmed in Fig.~\ref{fig:sigma-mx_bounds}, where the tightest constraints occur for Pop~III stars in dense DM environments and the modelling of helium scattering. An exciting result is the ability to constrain $\sigma$ below the neutrino floor for high DM densities!

\section{Conclusions}\label{sec:Conclusions}
In this paper, we extend the standard multiscatter capture formalism of~\cite{Bramante:2017,Ilie:2020Comment} to allow for the inclusion of various nuclear species in the capturing body, each with different masses and different scattering cross sections with DM particles. We then apply this formalism to Pop~III stars, and the effect of including the non-negligible amount of He inside those stars was evaluated. We find an enhancement of the capture rates of about one order of magnitude across all DM masses considered. This is solely  due to the $\sim 25\%$ He, which is much better at slowing down DM than the lighter H, and, additionally, has a larger cross section. We then proceed to impose projected bounds on DM parameters based on the potential discovery of Pop~III stars with JWST. We find, by imposing the sub-Eddington condition and assuming DM densities adiabatically enhanced in the high z host DM mini-halos, that Pop~III stars could be used to probe below the ``neutrino floor'' limiting direct detection experiments for $m_X\gtrsim 10^6$~GeV, when sufficiently high DM densities are considered. Even for the lowest DM density considered here, we find bounds on the spin independent DM-proton scattering cross section that are competitive with, or deeper than, those placed by the most sensitive experiment to-date, XENON1T \cite{Aprile:2018}. In the future, we plan to apply our formalism to other astrophysical objects that have been used as DM probes recently, and which are non single-component, such as exoplanets or white dwarfs.     

\section{Acknowledgements} CL thanks the financial support from Colgate University, via the Research Council student wage grant, and the Justus ’43 and Jayne Schlichting Student Research Funds.

\appendix

\section{DM scattering cross sections for Pop~III stars}\label{sec:SigmaAppendix}

The elastic scattering cross section between DM particles (X) and nuclei (a) is model-dependent and relies heavily on the DM-quark interaction strength as well as the distribution of quarks in the nucleons and nucleons in the nucleus. However, we can express generally a differential cross section as a function of recoil energy in the following way \cite{Jungman:1996}:

\be\label{eq:DiffCrossSection_Appendix}
d\sigma_a (E_R, v)= \sigma_0^a \frac{m_a}{2 \mu_a^2 v^2} F^2(E_R) d E_R,
\ee
where $\sigma_0^a$ is the ``standard" scattering cross section in the limit of 0 momentum transfer for nucleus a, $m_a$ is the mass of the nucleus, $\mu_a = \frac{m_a m_X}{m_a + m_X}$ is the DM-nucleus reduced mass, $v$ is the DM speed relative to the nucleus, $E_R$ is the recoil energy, and $F(E_R)$ is a nuclear form factor accounting for the non-null dimension of the nucleon, normalized such that $F(0) = 1$. The ``standard" cross section can be calculated from either spin-independent or spin-dependent interactions. In reality, the total cross section is a sum from the spin-independent and spin-dependent contributions, however, we assume SI interactions are dominant ($\sigma_0^a = \sigma_0^{SI-a}$) in this paper to study multi-component capture as helium has 0 spin. One could make the choice that SD interactions dominate, however, since only the protons in Pop~III stars would couple this way, that analysis is suited more towards the single-component analysis of Pop~III stars as done in \cite{Ilie:2020PopIIIa,Ilie:2020BNFa}. In order to get the total DM cross section, we must integrate the nuclear form factor over all possible recoil energies in the following way:
\be\label{eq:IntegrateSigma_Appendix}
\sigma_a = \sigma_0^{SI-a} \int^{E^{max}_R}_0 \frac{F^2(E_R)}{E^{max}_R }dE_R,
\ee
where $E^{max}_R = \frac{2 \mu_a^2 v^2}{m_a}$. 
We can then express the total DM-nucleon cross section in the following way:
\be\label{eq:SigmaIntegrated_Appendix}
\sigma_a = \sigma_0^{SI-a} \langle F^2(E_R)\rangle.
\ee
For the form factor, we use the Helm form factor \cite{Helm:1956,Duda:2007} and specifically adopt the conventions of \cite{Lewin:1995} in estimating necessary nucleus parameters. Thus, we define the helm factor in the following way: 
\be\label{eq:HelmFormFactor_Appendix}
     F^2(q) = \left(\frac{3 j_1(q R_1)}{q R_1} \right)^2 e^{-q^2 s^2},
\ee
 where $q = \sqrt{2 m_{a} E_R}$ is the momentum transfer of the collision, $j_1$ is the spherical bessel function of the first kind,
 $R_1$ is the effective nuclear radius, and $s$ is the nuclear skin thickness. Fitting Helm form factor parameters to muon spectroscopy
 data, \cite{Lewin:1995} finds:
 \be
     R_1 = \sqrt{c^2 + \frac{7}{3}(\pi a)^2 - 5 s^2},
\ee
where $c \se (1.23 A^{1/3} - 0.60)$ fm, with A being the atomic mass number,
$a \se 0.52$ fm, and $s \se 0.9$ fm. We note here that the form factor will always be 1 when considering interactions with protons. 

We now briefly estimate the error in taking the average of the form factor across all recoil energies. As an example, consider a $M_\star = 1000 M_\odot$ Pop~III star with surface escape velocity $v_{esc} = 5.4\times 10^8~\text{cm}~s^{-1}$. Taking the DM velocity as $v \simeq v_{esc}$ when it reaches the star, the maximum recoil energy of DM particles in the range $m_X = 10^2-10^{15}$ GeV off helium nuclei is $E_R^{max} \simeq 2.3-2.5$ MeV. The lack of significant variation is due to the limit $m_X \gg m_{He}$, reducing the maximum recoil energy to $E_R^{max} \approx 2 m_{He} v^2$, which is independent of DM mass. Evaluating the form factor in Eq.~\ref{eq:HelmFormFactor_Appendix} gives $F^2(E_R^{max}) \sim 0.5$, while the average given by the integral in Eq.~\ref{eq:IntegrateSigma_Appendix} is $\langle F^2(E_R)\rangle \sim 0.7$. Thus, averaging the form factor in this case is appropriate as the suppression is of order unity for all recoil energies.

The final piece required to estimate the total cross section for a given nucleus, $\sigma_a$, is the ``standard" cross section in the $q \to 0$ limit for SI interactions. For this, we have the following:
\be\label{eq:SigmaSI0Limit_Appendix}
\sigma_0^{SI-a} = A^2 \frac{\mu_a^2}{\mu_p^2} \sigma_0^{SI-p}  ,
\ee
where $\sigma_0^{SI-p}$ is the ``standard" DM-proton cross section in the $q \to 0$ limit, and $\mu_p = \frac{m_p m_X}{m_p+m_X}$ is the DM-proton reduced mass, under the assumption of equal DM coupling to neutrons and protons. In the limit of high DM mass relative to the target nucleus, valid for our analysis here ($m_X \gtrsim 100$ GeV$\gg m_p, m_{He} \sim$ GeV), one can simplify the DM-proton/nuclei reduced mass to: $\mu_p \simeq m_p$, $\mu_a \simeq m_a$. This means the factor $(\mu_a/\mu_p)^2$ in Eq.~(\ref{eq:SigmaSI0Limit_Appendix}) simplifies to $(m_a/m_p)^2$, which is simply $A^2$. With this in mind, one can simplify Eq.~(\ref{eq:SigmaSI0Limit_Appendix}) further to obtain:
\be\label{eq:SigmaSI_0Limit}
\sigma_0^{SI-a} \simeq A^4 \sigma_0^{SI-p}.
\ee
Combining this with Eq.~(\ref{eq:SigmaIntegrated_Appendix}) provides a way to calculate the DM-nucleus scattering cross section for Pop~III stars:
\be\label{eq:sigmaFINAL_Appendix}
\sigma_a \simeq A^4 \sigma_0^{SI-p} \langle F^2(E_R)\rangle .
\ee

In addition to suppressing the scattering cross section, the form factor also affects the distribution of momentum exchanges when DM scatters off a nucleus. In the multi-component multiscatter formalism, this information is encoded in the kinematic variable, $z$, whose average is taken to evaluate the capture rate. For protons, the result is simply $\langle z_p \rangle = 1/2$, as all possible momentum exchanges are evenly distributed. However, for an arbitrary nucleus $a$, a loss of coherence means that higher energy collisions are suppressed, and thus the average momentum exchange decreases. Following \cite{Bramante:2017}, we estimate the average of $z_a$ as: $\langle z_a\rangle \simeq \frac{\Eavg}{E_R^{max}}$. The average recoil energy, $\langle E_R\rangle$, can be estimated by an average of the recoil energy weighted with the form factor:
\be
\Eavg \simeq \frac{\int_0^{E_R^{max}} E_R F^2(E_R)~dE_R}{\int_0^{E_R^{max}} F^2(E_R)~dE_R}.
\ee

\section{Analytic expressions}\label{section:AnalyticExpressions}

In this section, analytic expressions for the total capture rate given by Eq.~(\ref{equation:CtotBasic}) are derived and used to find upper bounds on the DM parameters $\rho_X \times \sigma$. We start by introducing approximations for the probability function for DM undergoing N scatters. In the limits of $\tau \ll 1$ and $\tau \gg 1$, the function can be approximated by:
\begin{equation}
     p_N(\tau) \approx 
     \begin{cases}
         \frac{2 \tau^N}{N!(N+2)} + \mathcal{O}(\tau^{N+1}), & \text{if } \tau \ll 1\\
        \frac{2}{\tau^2}(N+1)\Theta(\tau - N), & \text{if } \tau \gg 1.\\
    \end{cases}
    \label{equation:PnApprox}
\end{equation}
To further simplify the expression in Eq.~(\ref{equation:CnSolved}), it is useful to explore the limiting regimes of the exponent by defining a new parameter, $R_{ij}$, such that:
\be
    R_{ij} \equiv \frac{3(v_{ij}^2 - v_{esc}^2)}{2 \Bar{v}^2}.
    \label{equation:Rv}
\ee
Expanding the exponent in Eq.~(\ref{equation:CnSolved}) gives:
\begin{equation}
    C_N \approx \sum_{i = 0}^{N}
    \begin{cases}
        \sqrt{\frac{2 \pi}{3}} n_X \frac{3 v_{esc}^2 + 2 \Bar{v}^2}{\Bar{v}} R^2 p_{i}(\tau_{A}) p_{j}(\tau_{B}) , & \text{if } R_{ij} \gg 1 \\
        \\
         \text{Const} p_i(\tau_A) p_j(\tau_B) \left(1 + i B_A\right) \left( 1 + j B_B\right) \left(j B_B + i B_A (1 + j B_B) \right), & \text{if } R_{ij} \ll 1,
    \end{cases}
    \label{equation:CnApprox}
\end{equation}
where we have introduced the simplifying notation $B_{A} \equiv \beta_+^{A}  \langle z_A \rangle$, $B_{B} \equiv \beta_+^{B} \langle z_{B} \rangle$, and $\text{Const} \equiv \sqrt{\frac{27 \pi}{2}} \frac{v_{esc}^4 n_X}{\vbar^3} R^2 $.~In the limit of $\tau \ll 1$, approximating the infinite sum in Eq.~(\ref{equation:CtotBasic}) is trivial as we are in the single scatter regime and can thus calculate the capture rate as $C_{tot} = C_1$. Recall that $\tau$ is defined such that it is the approximate average number of scatters a DM particle undergoes with a given component while traversing the object. Combining Eq.~(\ref{equation:CnApprox}) with Eq.~(\ref{equation:PnApprox}) under the assumption that $\tau_A, \tau_B \ll 1$, and noting that $C_{tot} \approx C_1$, gives the following analytic expressions for the total capture rate:
\be
    C_{tot} \approx 
    \begin{cases}
        \sqrt{\frac{8 \pi}{3}} R^2 n_X \frac{1}{\vbar} (\tau_A + \tau_B) \left(v_{esc}^2 + \frac{2}{3} \Bar{v}^2\right), & \text{if } \langle R_{ij}\rangle \gg 1\\
        \sqrt{6 \pi} R^2 n_X \frac{  v_{esc}^4}{\Bar{v}^3}\left[\tau_A B_A \left(1 + B_A \right) + \tau_{B} B_B \left(1 + B_B \right)\right], & \text{if } \langle R_{ij}\rangle  \ll 1,
    \end{cases}
    \label{equation:CtotApproxTauSmall}
\ee
where $\langle R_{ij} \rangle$ is the average of $R_{ij}$ across all scatters.
We now go on to derive analytic expressions in the $\tau_A,\tau_B \gg 1$ limit with an initial assumption that $\tau_B > \tau_A$ for the two limiting regimes of the exponent. For a single-component object, when $\tau \gg 1$, approximating the infinite sum in Eq.~(\ref{equation:CtotBasic}) can be done by truncating the sum at some $N_{cutoff} \sim \tau$. For a two-component object, a similar approach can be taken, however the sum is truncated at $\Ncut \sim \sum \tau$. 
\begin{figure}[bht]
    \centering
    \includegraphics[angle = 0, scale = 0.40]{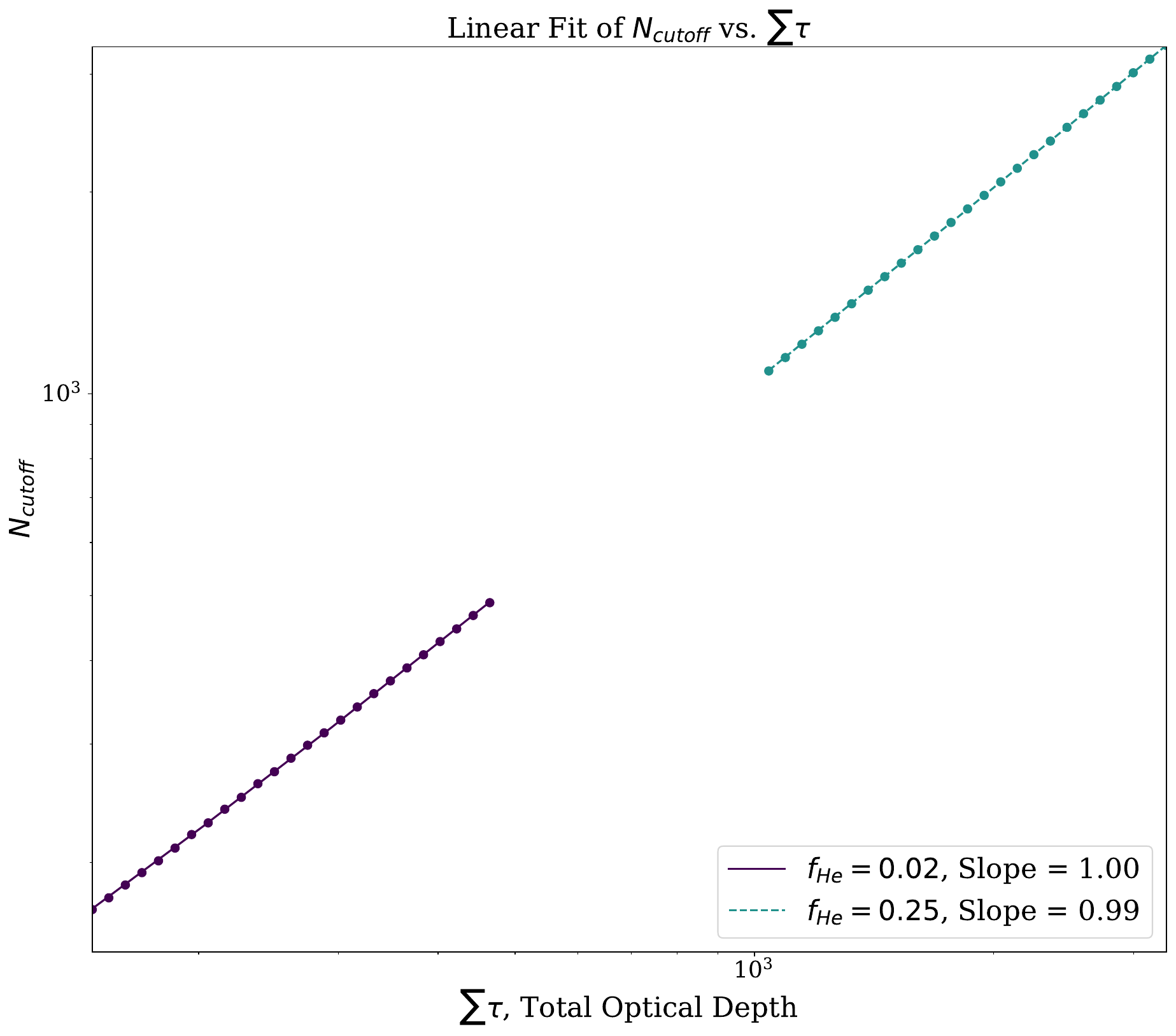}
    \caption{Total number of scatters for sum convergence $N_{cutoff}$, defined by Eqs.~(\ref{eq:Cutoff1}~-~\ref{eq:Cutoff2}), against the sum of optical depths $\sum \tau$, in a $M_\star = 100 M_\odot$ Pop~III star where $\sigma \sim 10^{-34} - 10^{-33}$ cm$^{2}$. The purple (blue) points represent the corresponding $N_{cutoff}$ when the capture rate converges for a given $\sum \tau$ in a $f_{He} = 0.02$ ($f_{He} = 0.25$) Pop~III star. The corresponding lines represent the linear fit to these points, which shows excellent agreement for a line with slope 1. This demonstrates the correlation between these two quantities and justifies the truncation of the infinite sum in Eq.~(\ref{equation:CtotBasic}) at $\sum \tau$ as a way to approximate the total capture rate in the limit of $\tauh, \tauhe \gg 1$.}
    \label{fig:Ncut_sumTau}
\end{figure}
To justify this assumption, we have plotted in Fig.~\ref{fig:Ncut_sumTau} $\Ncut$ from the convergence conditions in Eqs.~(\ref{eq:Cutoff1}~-~\ref{eq:Cutoff2}) against $\sum \tau$ for the case of a $M_\star = 100 M_\odot$Pop~III star, where we have taken $A = H$ and $B = He$.
In doing this, $\sigma$ is chosen arbitrarily in a range of $\sigma \sim 10^{-34}-10^{-33}$ cm$^2$ to get values for $\tauh, \tauhe \gg 1$ in the context of Pop~III stars. In addition, the fraction of helium in the star was artificially varied from the standard $f_{He}=0.25$ down to $f_{He} = 0.02$ in order to have optical depths of the same order of magnitude, i.e., $\tauh \sim \tauhe$. To see why this would be the case, refer to Eqs.~(\ref{equation:tauScale1}~-~\ref{equation:tauScale2}) which give scaling relationships for these quantities. These artificial selections are to verify the scaling relationship holds whether the optical depths are similar or not. As expected, a linear fit between these quantities has a slope of 1, and thus justifies summing to $\sum \tau$ as a valid approximation method for expressing $C_{tot}$ in the appropriate regime. In practice, because of the $\Theta(\tau - N)$ factor in Eq.~(\ref{equation:PnApprox}) for $\tau \gg 1$, which will appear in the expression for $p_A(\tau_A)$ and $p_B(\tau_B)$, the total capture rate can be approximated as:
\be
    C_{tot} \approx \sum_{N = 1}^{\tau_A} C_{N, low} + \sum_{N = \tau_A + 1}^{\tau_B} C_{N, mid} + \sum_{N = \tau_B + 1}^{\tau_B+\tau_A} C_{N, high},
    \label{eq:CtotApproxBigTau}
\ee
where we have defined $C_{N,low}$, $C_{N,mid}$, and $C_{N,high}$ in the following way for $R_{ij} \ll 1$: 
\be
    C_{N, low} \equiv D_1 \sum_{i = 0}^{N} (i+1)(j+1) \left(1 + i B_A \right) \left( 1 + j B_B \right) \left(j B_B + i B_A (1 + j B_B) \right) \Theta(\tau_A - i)\Theta(\tau_B - j),
    \label{eq:Clow}
\ee
\be
     C_{N, mid} \equiv D_1 \sum_{i = 0}^{\tau_A} (i+1)(j+1) \left(1 + i B_A \right) \left( 1 + j B_B \right) \left(j B_B + i B_A (1 + j B_B) \right) \Theta(\tau_A - i)\Theta(\tau_B - j),
     \label{eq:Cmid}
\ee
\be
     C_{N, high} \equiv D_1 \sum_{i = N_B}^{\tau_A} (i+1)(j+1) \left(1 + i B_A \right) \left( 1 + j B_B \right) \left(j B_B + i B_A (1 + j B_B) \right) \Theta(\tau_A - i)\Theta(\tau_B - j),
     \label{eq:Chigh}
\ee
where $D_1 \equiv \sqrt{216\pi} \frac{R^2 n_X}{\tau_{*}^2}\frac{v_{esc}^4}{\vbar^3}$, $\tau_{*} \equiv \tau_A \tau_B$, $N_B \equiv N-\tau_B$, $N = i + j$, and we have assumed $\tau_B > \tau_A$. Note that for $C_{N,low}$, $C_{N,mid}$, and $C_{N,high}$, the range of the sums is such that the factors $\Theta(\tau_A-i)\Theta(\tau_B-j)$ always evaluate to 1, making it simple to analytically evaluate them. Defining an analytic total capture rate in this way allows one to calculate up to the expected $\tau_A+\tau_B$ total average number of scatters with constituents in the object. Evaluating Eqs.~(\ref{eq:CtotApproxBigTau}~-~\ref{eq:Chigh}) yields the following final result in the $\tau_A, \tau_B \gg 1$ and $\langle R_{ij}\rangle \ll 1$ limit:
\begin{multline}\label{eq:CtotApproxBigTauFinal}
    C_{tot} \approx \frac{1}{144} D_1 (1+\tau_A)(2+\tau_A)(1+\tau_B)(2+\tau_B)\bigg[6 B_A\left[4+B_A\left(1+3\tau_A\right)\right]\\
    + 6 B_B\left[4+B_B\left(1+3\tau_B\right)\right] + B_A B_B \tau_A \tau_B \Big[48 + B_A B_B\left[1 + 3\left(\tau_A+\tau_B\right) + 9\tau_A\tau_B\right]\\
    + 8 B_A\left(1+3\tau_A\right)+ 8 B_B\left(1+3\tau_B\right)\Big]\bigg].
\end{multline}
Note the symmetry between $\tau_A$ and $\tau_B$, which explicitly demonstrates the possibility of using the same expression if $\tau_A > \tau_B$, and thus there is no loss of generality using the initial assumption $\tau_B > \tau_A$. A similar approach can be used to derive an expression for the total capture rate in the $\langle R_{ij}\rangle \gg 1$ and $\tau_A,\tau_B \gg 1$ limits. In doing this, Eq.~(\ref{eq:CtotApproxBigTau}) is still valid, however we must re-define $C_{N,low}$, $C_{N,mid}$, and $C_{N,high}$ in the following way:
\be
    C_{N, low} \equiv D_2 \sum_{i = 0}^{N} (i+1)(j+1)\Theta(\tau_A - i)\Theta(\tau_B - j),
    \label{eq:Clow_bigRv}
\ee
\be
     C_{N, mid} \equiv D_2 \sum_{i = 0}^{\tau_A} (i+1)(j+1)\Theta(\tau_A - i)\Theta(\tau_B - j),
     \label{eq:Cmid_bigRv}
\ee
\be
     C_{N, high} \equiv D_2 \sum_{i = N-\tau_B}^{\tau_A} (i+1)(j+1)\Theta(\tau_A - i)\Theta(\tau_B - j),
     \label{eq:Chigh_bigRv}
\ee
where $D_2 \equiv \sqrt{\frac{32 \pi}{3}} \frac{n_X}{\tau_*^2} \frac{3 v_{esc}^2 + 2 \Bar{v}^2}{\Bar{v}} R^2$. As with the case of $\langle R_{ij}\rangle \ll 1$, the sums in the equations above are over ranges such that the factor $\Theta(\tau_A - i)\Theta(\tau_B - j)$ always evaluates to unity. We can thus evaluate the sums analytically to obtain the following expression for the total capture rate in the $\tau_A, \tau_B \gg 1$ and $\langle R_{ij}\rangle \gg 1$ limits:
\begin{equation}
    C_{tot} \approx \frac{1}{4} D_2 \left(\tau_*^2 + 3 \tau_*\left( \tau _A+\tau _B \right)+9 \tau _*+2 \left( \tau _A^2 + \tau_B^2\right)+6 \left( \tau _A+ \tau _B\right)\right).
    \label{eq:Ctot_TauBig_RvBig}
\end{equation}

\begin{figure} [bht]
    \centering
    \includegraphics[angle = 0, scale = 0.45]{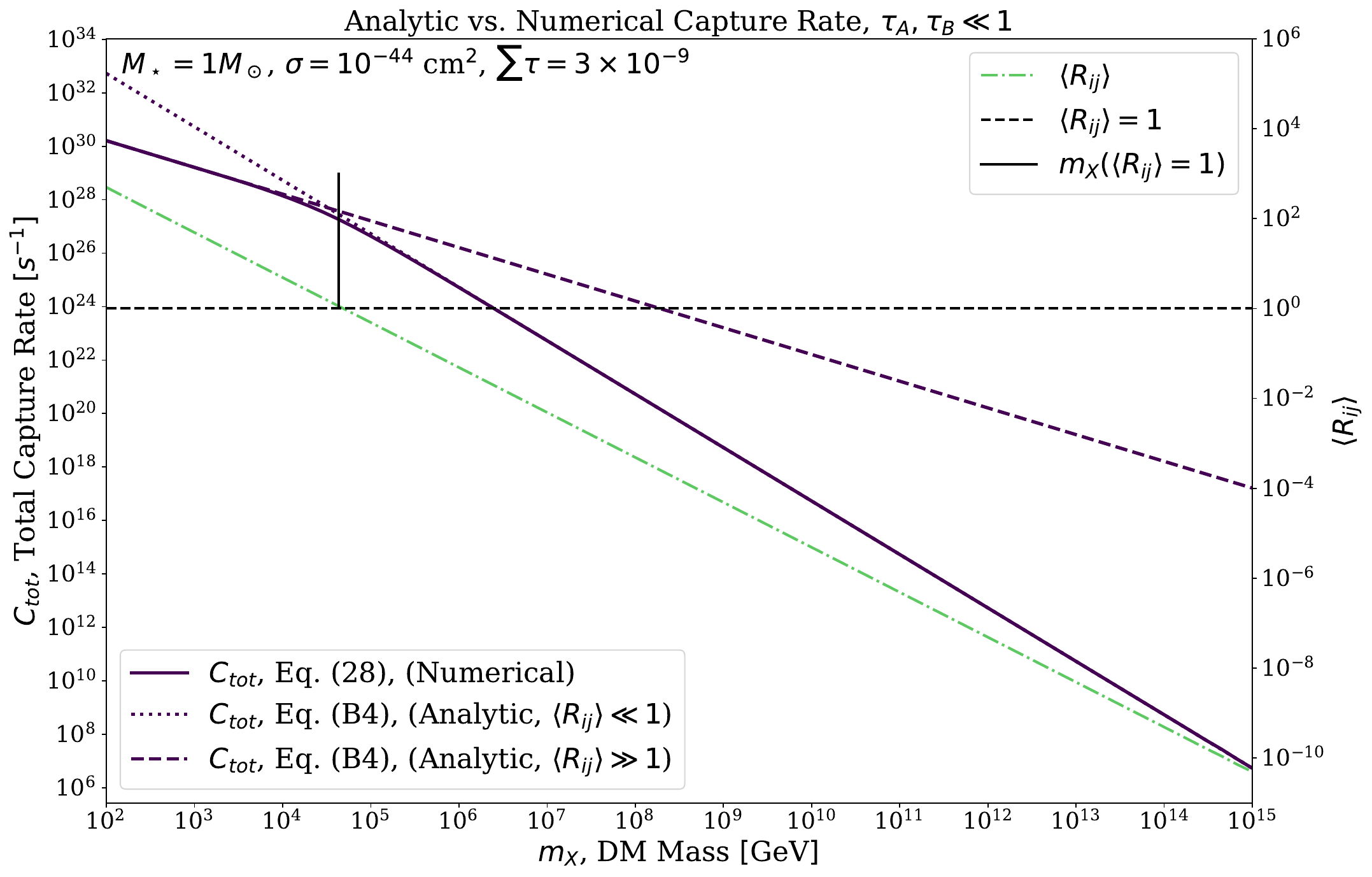}
    \caption{Full numerical and analytical capture rates using arbitrary Pop~III and DM parameters, demonstrating the validity of the analytic solutions in the $\tau_A, \tau_B \ll 1$ regime as well as the numerical accuracy of the multi-component multiscatter formalism using the truncation of the infinite sum described by Eqs.~(\ref{eq:Cutoff1}~-~\ref{eq:Cutoff2}). The solid purple line represents the full numerical solution while the dotted and dashed purple lines are the analytic solutions given by Eq.~(\ref{equation:CtotApproxTauSmall}) in the respective regimes. We also plot the average value of the exponential parameter $R_{ij}$ across all scatters for a given DM mass as a green dash-dotted line. This clearly demonstrates the transition in the behavior of the total capture rate as the average of this exponent across all scatters transitions from $\langle R_{ij}\rangle > 1$ to $\langle R_{ij}\rangle < 1$. Note that in this case, $\tau_A = \tau_H $ and $\tau_B = \tau_{He}$.}
    \label{fig:Ctot_NumvsAnalytic_smallTau}
\end{figure}

\begin{figure} [bht]
    \centering
    \includegraphics[angle = 0, scale = 0.45]{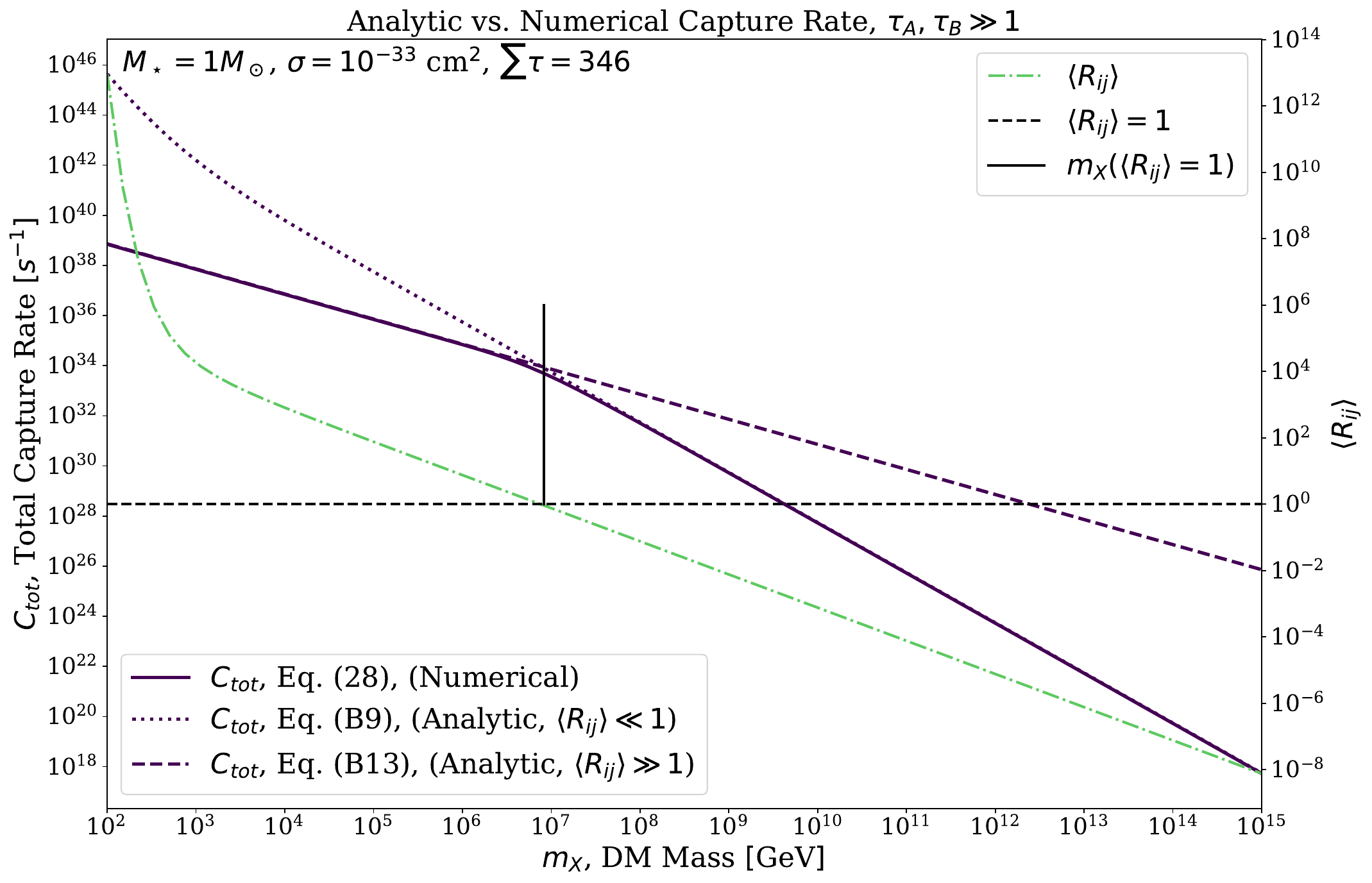}
    \caption{Full numerical and analytical capture rates using arbitrary Pop~III and DM parameters, demonstrating the validity of the analytic solutions in the $\tau_A, \tau_B \gg 1$ regime as well as the numerical accuracy of the multi-component multiscatter formalism using the truncation of the infinite sum described by Eqs.~(\ref{eq:Cutoff1}~-~\ref{eq:Cutoff2}). The solid purple line represents the full numerical solution while the dotted and dashed purple lines are the semi-analytic solutions given by Eqs.~(\ref{eq:CtotApproxBigTauFinal}) and (\ref{eq:Ctot_TauBig_RvBig}) in their respective regimes. We also plot the average value of the exponential parameter $R_{ij}$ across all scatters for a given DM mass as a green dash-dotted line. This clearly demonstrates the transition in the behavior of the total capture rate as the average of this exponent across all scatters transitions from $\langle R_{ij}\rangle > 1$ to $\langle R_{ij}\rangle < 1$. Note that in this case, $\tau_A = \tau_H $ and $\tau_B = \tau_{He}$.}
    \label{fig:Ctot_NumvsAnalytic_BigTau}
\end{figure}

In order to verify our analytic expressions, we compare them to the full numerical solutions described by the cutoff conditions in Eqs.~(\ref{eq:Cutoff1}~-~\ref{eq:Cutoff2}). Our results are found in Figs.~\ref{fig:Ctot_NumvsAnalytic_smallTau} and \ref{fig:Ctot_NumvsAnalytic_BigTau} which compare the numerical results to the analytic results in the $\tau_A, \tau_B \ll 1$ and $\tau_A, \tau_B \gg 1$ limits respectively. We note here that we chose Pop~III stellar parameters and DM parameters arbitrarily to verify the analytic expressions in the various limiting regimes. These plots also demonstrate the effect of the exponent $R_{ij}$ on the relationship between the capture rate and DM mass. The transition seen in the scaling of these two variables occurs when $\langle R_{ij} \rangle = 1$, as indicated in the plots. 


We now go on to use the expressions for the total capture rate to derive analytic expressions for constraints on $\rho_X \times \sigma$ from the observation of Pop~III stars due to the Eddington limit. As discussed in Sec.~\ref{sec:BoundsOnDM}, the Eddington limit can be used to bound DM properties as the observation of any Pop~III star implies a limit on DM heating from capture. By considering that a Pop~III star's total luminosity (DM + nuclear) must be less than the Eddington limit, we can use the following inequality to bound DM properties through Pop~III observation:
\be\label{eq:Ledd_LdmLnucApp}
L_{DM}(M_\star, \text{ DM params.}) \leq L_{edd}(M_\star) - L_{nuc}(M_\star),
\ee
where $L_{DM} = f C_{tot} m_X$. Hence, using the analytic equations for the total capture rates, we can solve for the important combination of DM parameters, $\rho_X \times \sigma$, and bound them. Although we have derived four distinct analytic expressions for the total capture rates based on the limits of $\tau_A,\tau_B$, and $\langle R_{ij}\rangle$, not all are valid for the capture rates calculated when saturating the inequality in Eq.~(\ref{eq:Ledd_LdmLnucApp}). It is straightforward to show that the regions that need to be considered for Pop~III star capture rates that saturate the Eddington limit are given by Eqs.~(\ref{equation:CtotApproxTauSmall}) and (\ref{eq:CtotApproxBigTauFinal}). We use another unique fact about Pop~III capture rates for DM masses $\gtrsim 10^2$ GeV to simplify these capture equations: $\beta_+^H, \beta_+^{He} \ll 1$, as $m_H, m_{He} \ll m_X$. We can thus rewrite the total capture rate for for a general two-component object capturing DM much heavier than the target nuclei, such as a Pop~III star, by taking the limit of $\beta_+ \gg 1$ in Eqs.~(\ref{equation:CtotApproxTauSmall}) and (\ref{eq:CtotApproxBigTauFinal}):
\be
    C_{tot} \approx 
    \begin{cases}
        \sqrt{\frac{8 \pi}{3}} R^2 n_X \frac{1}{\vbar} (\tau_H + \tau_{He}) \left(v_{esc}^2 + \frac{2}{3} \Bar{v}^2\right), & \text{if } \tau_H,\tau_{He} \ll 1 \text{ and }\langle R_{ij}\rangle \gg 1\\
        \sqrt{6 \pi} R^2 n_X \frac{  v_{esc}^4}{\Bar{v}^3}\left[B_H \tau_H  +  B_{He} \tau_{He}\right], & \text{if } \tau_H,\tau_{He} \ll 1 \text{ and }\langle R_{ij}\rangle \ll 1\\
        \sqrt{6\pi} R^2 n_X\frac{v_{esc}^4}{\vbar^3} \left[ B_H \tau _H + B_{He} \tau _{He}\right], & \text{if } \tau_H,\tau_{He} \gg 1 \text{ and }\langle R_{ij}\rangle \ll 1.
    \end{cases}
    \label{equation:CtotApprox_Pop3}
\ee
It is intriguing to note that in the limit of $\beta_+^H, \beta_+^{He} \ll 1$, the total capture rate is identical between the $\tauh, \tauhe \ll 1$ and $\tauh, \tauhe \gg 1$ limits when $\langle R_{ij}\rangle \ll 1$. This means that for Pop~III stellar capture in the regimes we consider, the limits of the analytic expressions depend only on the exponent $\langle R_{ij}\rangle$. It is then straightforward to find analytic bounds on the combination $\rho_X \times \sigma$ using Eqs.~(\ref{eq:Ledd_LdmLnucApp}) and (\ref{equation:CtotApprox_Pop3}), recalling that $\tau \sim \sigma$ and $n_X = \frac{\rho_X}{m_X}$. Solving for this combination of parameters gives: 
\be
    \rho_X \times \sigma \leq 
    \begin{cases}
        \frac{1}{\sqrt{6\pi}}
         \frac{\vbar}{(\frac{2}{3}\vbar^2+v_{esc}^2)} \frac{L_{edd} - L_{nuc}}{f\frac{4}{3} \pi R^3 (n_H + 256\langle F^2(E_R)\rangle n_{He})}, & \text{if } \langle R_{ij}\rangle \gg 1 \\
         \\
        \sqrt{\frac{2}{27\pi}}
         \frac{\vbar^3}{v_{esc}^4} \frac{L_{edd} - L_{nuc}}{f\frac{4}{3} \pi R^3 (B_H n_H + 256\langle F^2(E_R)\rangle B_{He} n_{He})}, & \text{if } \langle R_{ij}\rangle \ll 1,
    \end{cases}
    \label{equation:RhoSig}
\ee
where $n_{H}$ ($n_{He}$) is the average number density of hydrogen (helium) in the star. This expression can be used to place bounds on the combination of DM parameters $\rho_X \times \sigma$ for a Pop~III star of a given mass. Alternatively, one could place bounds on either parameter using assumptions on the other, as considered in Sec. \ref{sec:BoundsOnDM}.

\section{Generalized multi-component multiscattering formalism}\label{sec:ApMCMS}

In this section, a general formalism for calculating the DM capture rate in an object composed of $n$ different components is presented. The process is very similar to the one in Sec. \ref{sec:TwoCompformalism}, so a review of the details of the derivation given there is recommended. Start by considering a given astrophysical object comprised of $n$, evenly distributed components labeled as $\{I, II, III, ... , n\}$, with mass fractions given by $\sum_{m = 1}^{n} f_m = 1$. For each of these components, an optical depth is defined: $\{\tau_I, \tau_{II}, \tau_{III}, ... , \tau_n\}$, where $\tau_m \equiv 2 R_{obj} \sigma_m n_m$ for $m = I$ to $n$. As a DM particle traverses this object, there is a probability associated with collisions from $N = 1$ to $\infty$ with each component, given by:
\begin{equation}
    p_{N}(\tau_{m}) = 
    \begin{cases}
        \frac{2}{\tau_{m}^2} \left(N + 1 - \frac{\Gamma(N + 2, \tau_{m})}{N!}\right), & \text{if } \tau_{m} > 0\\
        \Theta(-N), & \text{if } \tau_{m} = 0,\\
    \end{cases}
\end{equation}
where $m$ ranges from $I$ to $n$ components. We now define an index for the number of scatters the DM particle has with each component as it traverses the star: $\{ \alpha, \beta, \gamma, ... , \omega\}$. Having collided with each component a number of times defined by the indices, the probability of being captured is given by:
\begin{equation} \label{eq:gN_general}
\begin{split}
g(w, \alpha, \beta, \gamma, ... , \omega) = \Theta &\left(v_{esc}\prod_{i = 1}^{\alpha}\left(1 - \langle z_{I}\rangle \beta_{+}^{I}\right)^{-\frac{1}{2}} \prod_{j = 1}^{\beta}\left(1 - \langle z_{II}\rangle \beta_{+}^{II}\right)^{-\frac{1}{2}}\times  \ldots \right.\\
&\left. \quad \times \prod_{y = 1}^{\omega}\left(1 - \langle z_{n}\rangle \beta_{+}^{n}\right)^{-\frac{1}{2}} - w\right),
\end{split}
\end{equation}
where $w(r)^2 = u^2 + v_{esc}(r)^2$ and $u$ is the DM velocity far from the star. We then define the partial capture rate, i.e., the capture rate after collisions given by $\{ \alpha, \beta, \gamma, ... , \omega\}$ collisions with $\{I, II, III, ... , n\}$ components, as:
\begin{equation}
    C(\alpha, \beta, \gamma, ... , \omega) = \pi R^{2} p_{\alpha}(\tau_{I}) p_{\beta}(\tau_{II}) \times ... \times p_{\omega}(\tau_{n}) \int_{v_{esc}}^{\infty} dw \frac{f(u)}{u^2} w^3  g(w, \alpha, \beta, \gamma, ... , \omega).
    \label{equation:Cpartial_GeneralMCMS}
\end{equation}
To calculate the total capture rate, one must then carry out $n$ sums over the partial capture rate in the following way:
\begin{equation}\label{eq:Ctot_GeneralMCMS}
    C_{tot} = \sum_{\alpha = 1}^{\infty} \sum_{\beta = 1}^{\infty} ... \sum_{\omega = 1}^{\infty} C(\alpha, \beta, \gamma, ... , \omega).
\end{equation}
In practice, as demonstrated throughout this paper with a two-component system, these sums converge at a given number of collisions that is dependent on $\sum_i^n \tau_i$, and so it is useful to derive a way to calculate the number of terms one would have to sum to find the capture rate up to $N$ collisions in an $n$-component object. Another way to phrase this question is, given a DM particle collides $N$ different times in an object with $n$ components, how many different ways can this happen? To answer this, we point to the following equation:
\begin{equation}
    T(N, n) = \frac{1}{2}n^2(N-1) + \frac{1}{2}n(3-N),
\end{equation}
where $T$ is the number of terms to sum. This quadratic relationship clearly demonstrates the computational price incurred by increasing the number of components to be considered for collisions and is pertinent to keep in mind when calculating capture rates in multi-component objects.

\section{Verification of numerical convergence}\label{Sec:NumericalTest}
In this section, we address issues relating to the feasibility of calculating multi-component capture rates and the validity of the numerical convergence criteria established in Eqs.~(\ref{eq:Cutoff1}~-~\ref{eq:Cutoff2}). As shown in the previous section, there is a significant cost incurred when the number of components considered for capture is increased. Recall that the total capture rate in a two-component context can be calculated as an infinite sum of partial capture rates given by the following equation:
\begin{equation}
    C_{tot} = \sum_{N = 1}^{\infty} C_N,
\end{equation}
where the partial capture rate $C_N$, is:
\begin{equation}
    C_{N} = \sum_{i = 0}^{N}\left[\pi R^{2} p_{i}(\tau_{H}) p_{j}(\tau_{He}) \int_{v_{esc}}^{\infty} dw \frac{f(u)}{u^2} w^3 g_{ij}(w)\right].
\end{equation}
However, for the purposes of increasing computational efficiency through parallelized algorithms, one can reformulate the idea of the total capture rate in a two-component object as a double sum to infinity of a partial capture rate defined in the following way: 
\begin{equation}
    C_{ij} = \pi R^{2} p_{i}(\tau_{H}) p_{j}(\tau_{He}) \int_{v_{esc}}^{\infty} dw \frac{f(u)}{u^2} w^3 g_{ij}(w),
    \label{equation:Cij}
\end{equation}
where the total capture rate would then be given by:
\begin{equation}
    C_{tot} = \sum_{i = 1}^{\infty} \sum_{j = 1}^{\infty} C_{ij}.
\end{equation}
In practice, these would be summed to cutoffs dependent on their respective optical depths in the following way:
\begin{equation}\label{eq:Ctot_CijDoubleSum}
    C_{tot} \approx \sum_{i = 1}^{N_{cut,H}} \sum_{j = 1}^{N_{cut,He}} C_{ij},
\end{equation}
where $N_{cut,H} \sim \tauh$ and $N_{cut,He} \sim \tauhe$. 

\begin{figure}[!htb]
\centering
\includegraphics[angle = 0, scale = 0.6]{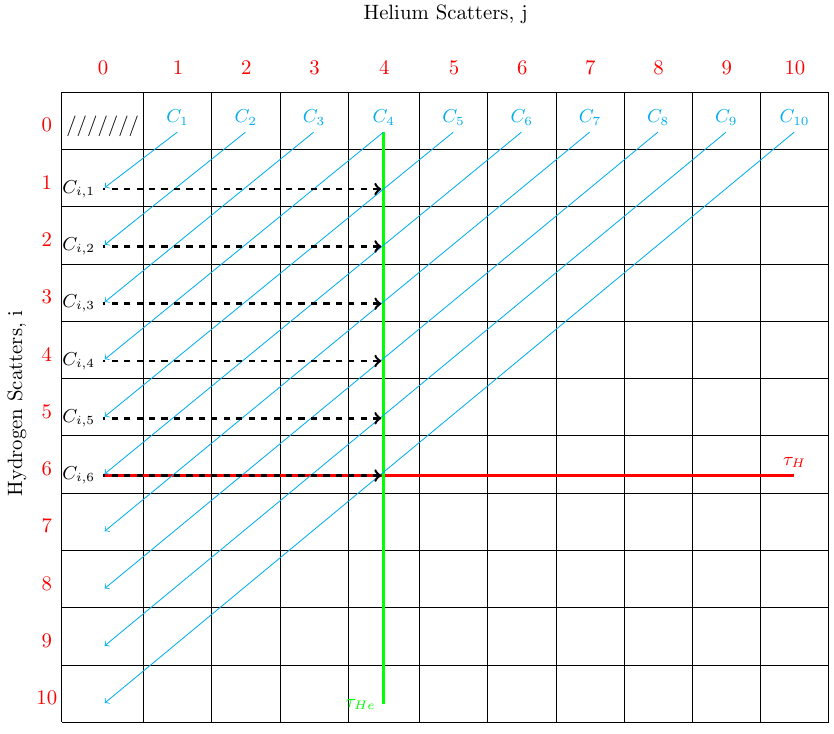}
\caption{Schematic diagram demonstrating the two equivalent ways of calculating the total DM capture rate in Pop~III stars. Each cell in the diagram, $(i,j)$, represents the partial capture rate, $C_{ij}$, for $i$ scatters with hydrogen and $j$ scatters with helium nuclei. The dashed lines represent the strategy of obtaining the total capture rate given by Eq.~(\ref{eq:Ctot_CijDoubleSum}), where a double sum is performed over partial capture rates $C_{ij}$. The solid, cyan lines represents the strategy outline by Eq.~(\ref{eq:Ctot_toNcut}), where partial capture rates are defined along the diagonals as shown. The red (green) line represents the optical depth for hydrogen (helium). One thing to note here is that while the method of summing along the diagonals seems like it would lead to a higher total capture rate, as more cells are included, when $i > \tauh$ and/or $j >\tauhe$, the capture rate is significantly suppressed (See Fig.~\ref{fig:Cij_Heatmap}).}\label{fig:Cij_schematic}
\end{figure}

\begin{figure} [bht]
    \centering
    \includegraphics[angle = 0, scale = 0.4]{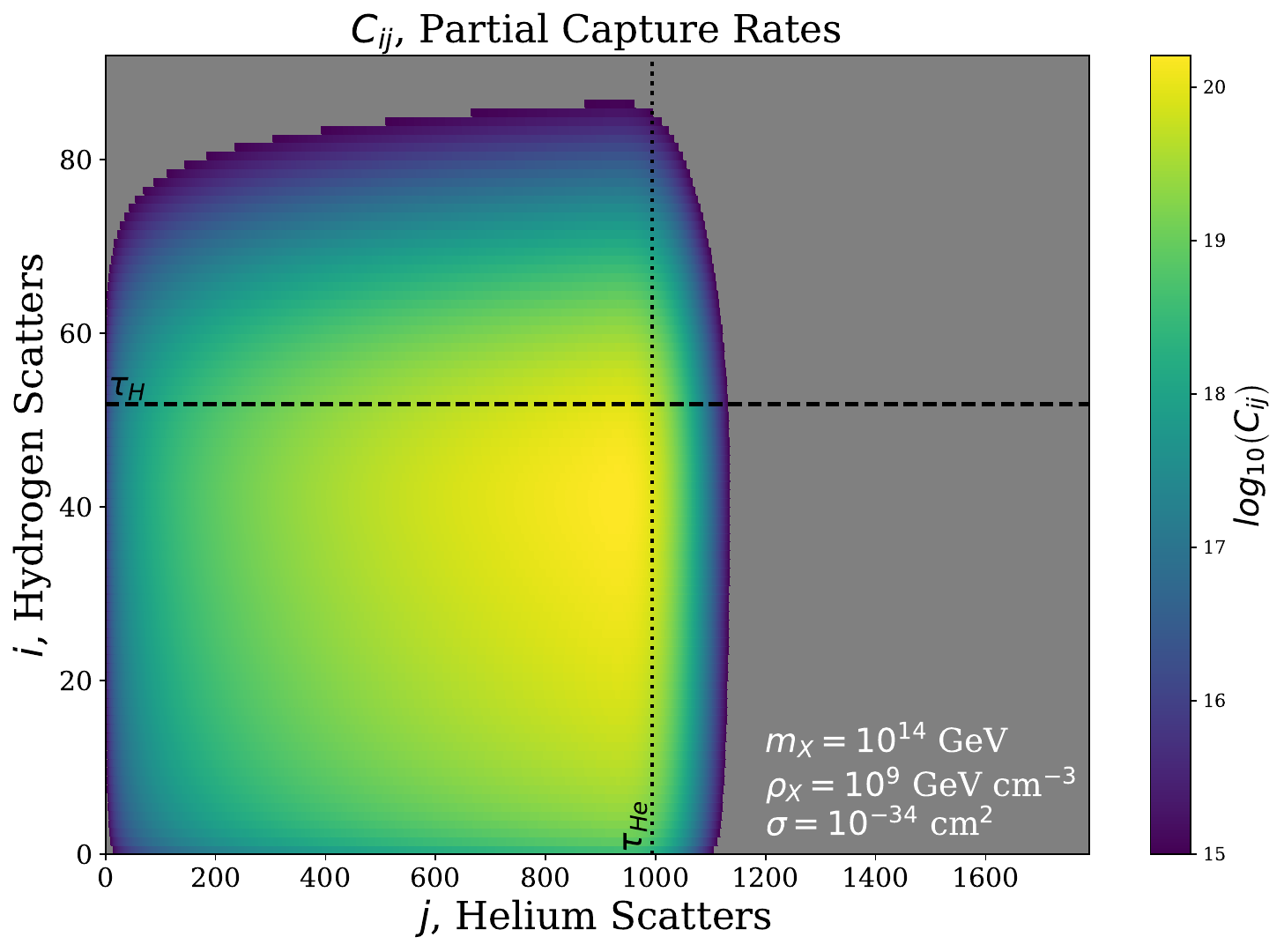}
    \caption{Partial capture rates defined by Eq.~(\ref{equation:Cij}) for a $M_\star = 100 M_\odot$ Pop~III star for a given $i$ number of scatters with hydrogen nuclei and $j$ scatters with helium nuclei. The brighter colors represent higher capture rates and the dashed (dotted) line represents $\tauh$ ($\tauhe$). As expected, the partial capture rates are peaked around $i\sim \tauh$ and $j\sim \tauhe$ and drop off rapidly when these values are exceeded. This is because the optical depth $\tau$ is defined to be the average number of scatters a DM particle will undergo with a given nucleus while traversing the object. This provides justification for the truncation of each sum in Eq.~(\ref{eq:Ctot_CijDoubleSum}) at $\sim \tau$.}
    \label{fig:Cij_Heatmap}
\end{figure}

Fig.~\ref{fig:Cij_schematic} schematically depicts the two equivalent ways of calculating the total capture rates we have discussed, i.e., summing $C_N$ and double-summing $C_{ij}$. The advantage of reformulating the computation in this way is that implementing a parallelized algorithm is much simpler when summing horizontally in Fig.~\ref{fig:Cij_schematic}, as each row can simply be summed on separate cores and recombined afterwards. In Fig.~\ref{fig:Cij_Heatmap}, we plot an array of $C_{ij}$ as a heatmap to give clarity on how the partial capture rates become suppressed after surpassing the optical depth along the hydrogen and helium scattering axes.
This serves as verification for summing up to a cutoff that depends on each optical depth when considering $C_{ij}$ partial capture rates per Eq.~(\ref{eq:Ctot_CijDoubleSum}), and summing up to a cutoff depending on the sum of optical depths, as per Eq.~(\ref{eq:Ctot_toNcut}).

\begin{figure} [!h]
    \centering
    \includegraphics[angle = 0, scale = 0.4]{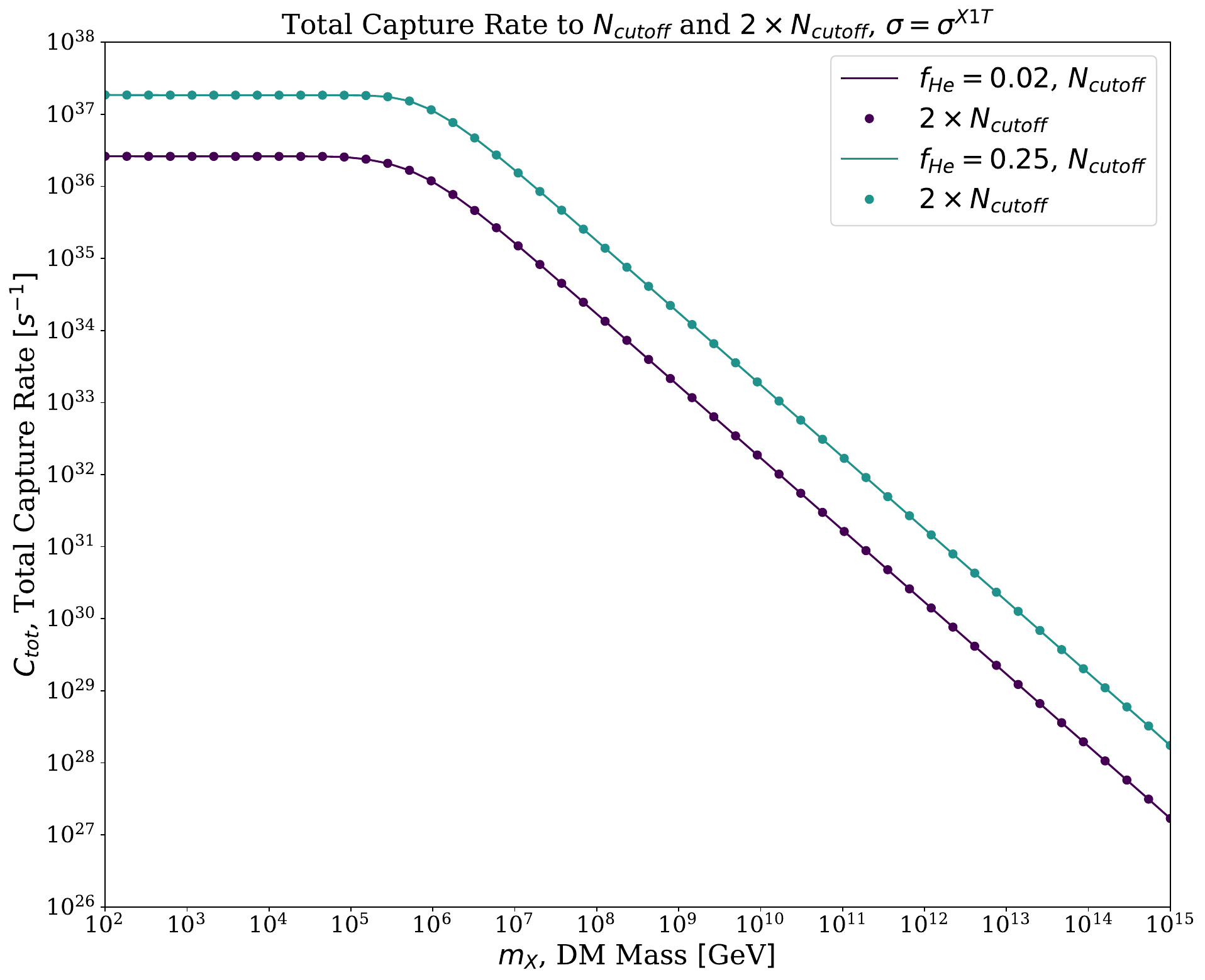}
\caption{Total capture rate in a $M_\star = 100 M_\odot$ Pop~III star where the infinite sum in Eq.~(\ref{equation:CtotBasic}) is truncated at $N_{cutoff}$ as defined by Eqs.~(\ref{eq:Cutoff1}~-~\ref{eq:Cutoff2}) and $2\times N_{cutoff}$. The solid lines represent the truncation up to $N_{cutoff}$ while the points represent up to $2 \times N_{cutoff}$. The color purple (blue) represents $f_{He}= 0.02$ ($f_{He}=0.25$). Note the perfect agreement between the sums to $N_{cutoff}$ and $2\times N_{cutoff}$, indicating that the sums have converged by $N_{cutoff}$ and thus the convergence criteria are valid.}
    \label{fig:Ctot_ConvergenceTest}
\end{figure}

Cutoff criteria for sum convergence were imposed to calculate the total capture rate based on the fact that the capture rate falls rapidly when the number of scatters considered surpasses the average number of scatters, defined by a sum of the average number of scatters with each component, $\sum \tau = \tauh + \tauhe$.
The total capture rate is then approximated by a sum to $N_{cutoff}$:
\be
    C_{tot} \approx \sum_{N = 1}^{N_{cutoff}} C_N.
\ee
Initial verification of this convergence criteria can be found in Appendix \ref{section:AnalyticExpressions}, where analytic expressions for the capture rates in limiting regimes are derived and compared to the numerical solution. To further verify that the sums have indeed converged, total capture rates for a Pop~III star of mass $M_\star = 100 M_\odot$ up to $N_{cutoff}$ and $2 N_{cutoff}$ are calculated. This is done for two cases, one where the fraction of helium in the star is $\sim 25\%$ and another where the fraction of helium is artificially imposed to $\sim 2\%$, to test the criteria for realistic optical depths (where $f_{He} = 0.25$ implies $\tauhe \sim 10^2~\tauh$) and optical depths of a similar order of magnitude (where $f_{He} = 0.02$ implies $\tauhe \sim \tauh$). The results can be found in Fig.~\ref{fig:Ctot_ConvergenceTest}, which show perfect agreement between summing up to the cutoff value and twice this value, demonstrating that the sums have indeed converged by $N_{cutoff}$. 

\bibliography{RefsDM}
\end{document}